\documentclass[10pt,5p,twocolumn,authoryear]{elsarticle}

\usepackage{amsmath}
\usepackage[hyphens]{url}
\usepackage{float}

\usepackage{graphicx}
\graphicspath{{.}}
\usepackage{float}

\newcommand\blfootnote[1]{%
  {
  \renewcommand\thefootnote{}\footnote{#1}%
  \addtocounter{footnote}{-1}%
  }
}

\newcommand{\mrev}[2]{#2}
\newcommand{\mrevi}[2]{#2}

\newcommand{\LpSsimSuppl}[1]{S1#1}
\newcommand{\OtherMethodsBMIa}[1]{S2#1}
\newcommand{\OtherMethodsBMIb}[1]{S3#1}
\newcommand{\noncvxRPCAHCPBMIRand}[1]{S4#1}
\newcommand{\FigOutlierODF}[1]{S5#1}

\newcommand{\beginsupplement}{%
        \setcounter{table}{0}
        \renewcommand{\thetable}{S\arabic{table}}%
        \setcounter{figure}{0}
        \renewcommand{\thefigure}{S\arabic{figure}}%
     }
\newcommand{\noncvxRPCAHCP}{7}

\begin{document}

\title{{Low Rank plus Sparse Decomposition of \mrevi{R2.2}{ODFs} for Improved Detection of Group-level Differences and Variable Correlations in White Matter}}

\author[CAIR,CBI]{Steven H. Baete\corref{cor1}}
\ead{steven.baete@nyumc.org}
\author[CAIR,CBI,Psych]{Jingyun Chen}
\author[CAIR,CBI]{Ying-Chia Lin}
\author[CAIR,CBI]{Xiuyuan Wang}
\author[CAIR,CBI]{Ricardo Otazo}
\author[CAIR,CBI]{Fernando E. Boada}
\address[CAIR]{Center for Advanced Imaging Innovation and Research (CAI$^2$R), NYU School of Medicine, 660 First Ave 4th Floor, New York, NY 10016, USA}
\address[CBI]{Center for Biomedical Imaging, Dept. of Radiology, NYU School of Medicine, 660 First Ave 4th Floor, New York, NY 10016, USA}
\address[Psych]{Dept. of Psychiatry, NYU School of Medicine, One Park Avenue, New York, NY 10016, USA}
\cortext[cor1]{Corresponding author at: NYU School of Medicine, Dept. of Radiology, 660 First Ave 4th Floor, New York, NY 10016, USA}

\begin{abstract} 
A novel approach is presented for group statistical analysis of diffusion weighted MRI datasets through voxelwise Orientation Distribution Functions (ODF).

Recent advances in MRI acquisition make it possible to use high quality diffusion weighted protocols (multi-shell, large number of gradient directions) for routine in vivo study of white matter architecture. The dimensionality of these data sets is however often reduced to simplify statistical analysis. While these approaches may detect large group differences, they do not fully capitalize on all acquired image volumes. Incorporation of all available diffusion information in the analysis however risks biasing the outcome by outliers.

Here we propose a statistical analysis method operating on the ODF, either the diffusion ODF or fiber ODF. To avoid outlier bias and reliably detect voxelwise group differences and correlations with demographic or behavioral variables, we apply the Low-Rank plus Sparse ($L+S$) matrix decomposition on the voxelwise ODFs \mrevi{R2.3}{which} separates the sparse individual variability in the sparse matrix $S$ whilst recovering the essential ODF features in the low-rank matrix $L$.


We demonstrate the performance of this ODF $L+S$ approach by replicating the established negative association between global white matter integrity and physical obesity in the Human Connectome dataset. \mrevi{R2.4}{The volume of positive findings ($p < 0.01$, 227cm$^3$) agrees with and expands on the volume found by TBSS (17cm$^3$), Connectivity based fixel enhancement (15cm$^3$) and Connectometry (212cm$^3$).} In the same dataset we further localize the correlations of brain structure with neurocognitive measures such as fluid intelligence and episodic memory.

The presented ODF $L+S$ approach will aid in the full utilization of all acquired diffusion weightings leading to the detection of smaller group differences in clinically relevant settings as well as in neuroscience applications.\\


\noindent{\textcopyright 2018}. This manuscript version is made available under the CC-BY-NC-ND 4.0 license \url{http://creativecommons.org/licenses/by-nc-nd/4.0/}
\end{abstract}

\maketitle

\blfootnote{\noindent{\it Abbreviations\/}:
BMI, Body Mass Index;
DSI, Diffusion Spectrum Imaging;
DWI, Diffusion Weighted Imaging;
DTI, Diffusion Tensor Imaging;
fODF, fiber Orientation Distribution Function;
FWE, Family Wise Errors;
GLM, General Linear Models;
GQI, Generalized Q-Space Sampling;
HARDI, High Angular Resolution Diffusion Imaging;
HCP, Human Connectome Protocol;
JSD, Jensen-Shannon Divergence;
ODF, Orientation Distribution Function;
noncvxRPCA, Robust PCA via Nonconvex Rank Approximation algorithm;
PC, Principal Component;
PCA, Principal Component Analysis;
QA, Quantitative Anisotropy;
RPCA, Robus Principal Component Analysis;
SNR, Signal to Noise Ratio;
TBSS, Tract-Based Spatial Statistics;
TFCE, Threshold-Free Cluster Enhancement;
VBA, Voxel-Based Analysis.}

\section{Introduction}

Diffusion weighted magnetic resonance imaging (DWI MRI) samples the diffusive displacement of water and its interactions with cellular structures such as axon membranes in \textit{in vivo} white matter \citep{Callaghan1993,Basser1996}. By encoding the anisotropic tissue micro-structure, DWI MRI provides insight in the complex white matter tract architecture \citep{Wedeen2012,Fernandez-Miranda2012}. The crossing fibers translate in each voxel to Orientation Distribution Functions (ODF \citep{Callaghan1993}), captured by detailed High Angular Resolution Diffusion Imaging (HARDI, \citep{Tuch2002}) methods such as Diffusion Spectrum Imaging (DSI \citep{Callaghan1993,Wedeen2005,Wedeen2012,Reese2009}) and Q-ball imaging \citep{Tuch2004}.\par


The long acquisition times, imposed by the large number of q-space samples needed to accomplish sufficient angular resolution, have long hindered widespread adoption of HARDI datasets in group studies \citep{Kuo2008, Setsompop2012b}. Recent developments in simultaneous multi-slice or multiband techniques \citep{Setsompop2012, Blaimer2013} and sequence design \citep{Baete2015ISMRMDSIMB, Baete2015RDSI, Baete2017} have however led to data acquisition times that, for the first time, make HARDI a routine viable and practical tool for clinical applications and neuroscience research. This evolution has highlighted the need for robust methodologies for statistical analysis of group ODF datasets.\par

A number of methods have been proposed to identify and study differences in the diffusion signals of groups of subjects. Diffusion-specific Voxel-Based analysis (VBA) methods register quantitative diffusion measures for the whole brain \citep{Whitcher2007} or project them on a tract skeleton (Tract-Based Spatial Statistics, TBSS \citep{Smith2006,Jbabdi2010}) or surface \citep{Zhang2010}. Most of these approaches are based on information gained from Diffusion Tensor Imaging (DTI, \citep{Basser1996}), an incomplete representation of the complex intravoxel crossings in the human brain \citep{Jeurissen2012}. This incomplete representation is partially mitigated by an extension of the TBSS-method accommodating two crossing fibers \citep{Jbabdi2010}. Nevertheless, the focus of these methods on DTI makes them ill-suited to fully exploit the much higher dimensionality of ODFs. In addition, the projection based methods suffer from inaccurate tract representations and projections \citep{Bach2014,Raffelt2015}.\par

Other methods use tractography results to identify structurally connected fiber populations globally \citep{Jahanshad2015,Chen2015HBM,Raffelt2012,Raffelt2015} or locally \citep{Yeh2016}. The resulting connectivity matrices can then be used directly for statistical tests \citep{Jahanshad2015,Chen2015HBM,Mitra2016} or the tractograms can inform tract-specific smoothing \citep{Raffelt2012,Raffelt2015} and enhancement of statistical maps along the tracts \citep{Raffelt2012,Raffelt2015,Yeh2016} using Threshold-Free Cluster Enhancement approaches (TFCE, \citep{Smith2009}). Whilst these tractography based methods are powerful, they suffer from problems related to imperfections in the tractography \citep{Jones2013,Reveley2015,Thomas2014}, some limit the identified fiber directions to a predefined template \citep{Yeh2016} and they generally miss more subtle differences in diffusion patterns conveyed in the ODF.\par

The most promising methods for group difference identification in diffusion MRI studies capitalize on the full dimensionality of the rich information contained in ODFs registered to a common atlas. Early work used voxelwise whole brain multivariate statistics on the coefficients of spherical harmonic representations of ODFs \citep{Lepore2010MICCAI}. The first approach to mine the high dimensionality of the whole ODF rather than a representation, applied Principal Component Analysis (PCA) to identify the defining ODF features in each voxel in a whole brain group analysis \citep{Chen2015HBM}. Statistical analysis of the weights of the Principal Components, the PC-scores, then informs the significance of group differences \citep{Chen2015HBM}. However, PCA is sensitive to outliers and can be easily corrupted by the individual variability of subjects \citep{Zhou2014,Lin2016}, reducing the power of the statistical test.\par

Here, we extend the previous work by isolating the essential ODF features in each voxel which are common/different within/between subject cohorts from the individual subject variability. This is achieved by replacing the PCA of ODF distributions by a Low-Rank plus Sparse ($L+S$) Matrix Decomposition \citep{Candes2011,Chandrasekaran2011,Otazo2015}. The $L+S$ decomposition, also referred to as Robust PCA (RPCA), separates the sparse individual variability in the sparse matrix $S$ whilst recovering the essential ODF features in the low-rank matrix $L$. Subsequently, statistical tests can focus on the defining ODF features in $L$, thus increasing the detectability of group differences and correlations in diffusion datasets. This is then extended to a whole brain analysis using Threshold-Free Cluster Enhancement (TFCE, \citep{Smith2009}) to correct for Family Wise Errors (FWE). Although we will apply this technique here to the diffusion ODF, as derived from Q-Ball imaging \citep{Tuch2004}, DSI \citep{Callaghan1993}, or Generalized Q-Space Sampling (GQI,\citep{Yeh2010,Yeh2011}), it is also applicable to the fiber ODF (fODF) obtained by spherical deconvolution \citep{Tournier2004}.\par

\begin{figure}[tb]
    \begin{center}
    \includegraphics[width=0.45\textwidth,trim=0 0 0 0, clip]{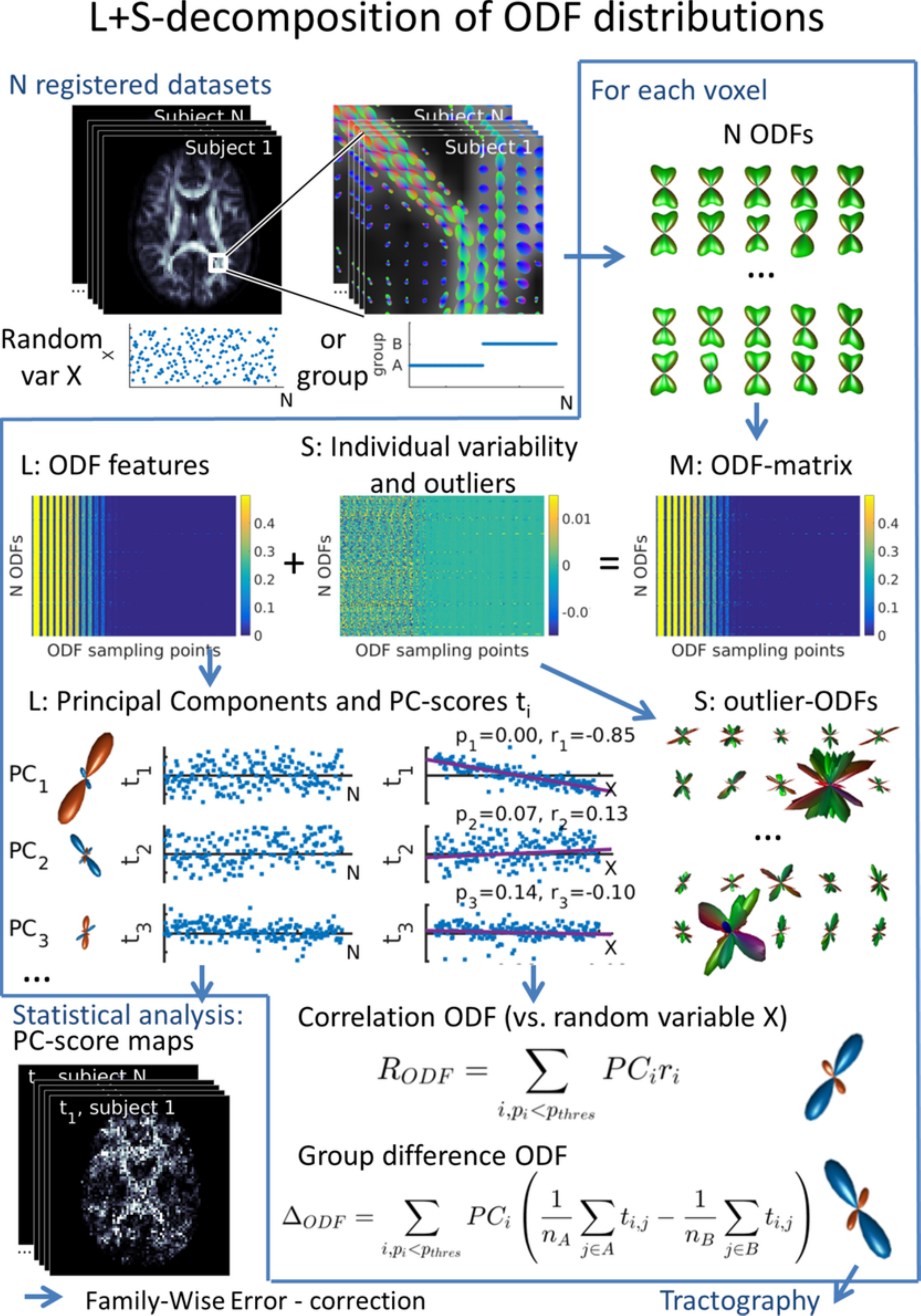}
        \caption{Differences between or correlations in groups of ODFs, taken from registered voxels, are identified by reorganizing the ODF-values in an ODF-matrix $M$. Subsequently, the $L+S$ decomposition isolates the features of $M$ common/different between groups in $L$ and splits the individual variability and outliers away in $S$. Significance of ODF differences is assessed by a statistical analysis of the Principal Component scores ($t$) of $L$ ($p_L$). Difference ODFs $\Delta_{ODF}$ can be calculated from the significant group differences while correlation ODFs $R_{ODF}$ can be calculated from significant correlations with a random variable.\label{LpSScheme}}
    \end{center}
\end{figure}

The $L+S$ matrix decomposition is ideally suited for the isolation of the essential low-rank ODF features in $L$. Indeed, the exact recovery of the $L$ and $S$ components has been mathematically proven under limited restrictions of rank and sparsity \citep{Candes2011,Chandrasekaran2011}. These mathematical properties are exploited in, amongst others, image alignment \citep{Peng2012}, denoising and background extraction in video \citep{Bouwmans2014,Lin2016}, segmentation of images and video \citep{Bouwmans2014,Lin2016}, reconstruction of diffusion MRI \citep{Gao2013ISMRM0610}, dynamic CT \citep{Gao2011b} and MRI \citep{Otazo2015} images, and filtering of fMRI datasets \citep{Otazo2015HBM1690}.\par

The Principal Components of $L$ would identify in each voxel the essential ODF features which can be used to calculate group difference $\Delta_{ODF}$ and correlation $R_{ODF}$ ODFs. $\Delta_{ODF}$ and $R_{ODF}$ visualize significant group differences and correlations and serve as input for tractography similar to the local tractography visualization approach used in \citep{Yeh2016}.\par

In this work, we introduce the use of $L+S$ matrix decomposition for examining ODF group differences and correlations with biological measurements in HARDI datasets. We establish the applicability and feasibility in theoretical analysis and computer simulations. Subsequently, we demonstrate the approach by confirming the well-established negative association between global white matter integrity and physical obesity \citep{Gianaros2013, Mueller2011, Stanek2011, Verstynen2013, Verstynen2012, Yeh2016} in a cohort of healthy Human Connectome Project volunteers. Finally, in the same cohort we explore white matter areas correlated to motor functioning, language and vocabulary comprehension and decoding, episodic memory and fluid intelligence \citep{Smith2015, Powell2017}.\par

\section{Theory}

The ODFs in each voxel of a set of registered whole brain diffusion datasets can be expected to be highly correlated within that voxel (Fig. \ref{LpSScheme}). Although subject subgroup differences can arise, one can assume that all ODFs will be drawn from a lower-dimensional subspace. This means that the ODF features which are common between or within subject groups will be captured in that low-dimensional subspace if we, for a moment, ignore the sparse individual variability. The low-dimension assumption of the ODFs can be translated in an assumption of low rank \citep{Zhou2014, Lin2016} where the rank of a matrix is the number of linearly independent rows or columns that define a basis set to represent the matrix. Hence, once we identify the low-rank subspace of the ODFs, we can easily identify the ODF features different between subject subgroups.\par

Earlier work used PCA to identify the key ODF features \citep{Chen2015HBM}. While this method does share the philosophy of determining the low-rank basis of the ODFs, PCA is easily corrupted by gross errors due to its assumption of independently and identically distributed Gaussian noise \citep{Zhou2014}. If the individual variability is non-Gaussian and strong, even a few outliers can make PCA fail \citep{Lin2016}.\par

The exact recovery of both the low-rank and sparse components of matrices \citep{Candes2011,Chandrasekaran2011} has been of great interest in a number of applications \citep{Peng2012,Bouwmans2014,Lin2016,Gao2013ISMRM0610,Gao2011b,Otazo2015,Otazo2015HBM1690}. Separating these components allows focus on either the common features or the dynamic aspects of datasets, respectively the low-rank $L$ and sparse $S$ submatrices. The $L+S$ matrix decomposition is commonly expressed as
\begin{align}
  \text{minimize} \,\,||L||_* + \lambda ||S||_1 \label{LpSeq}\\
  \text{subject to} \,\,L+S = M \nonumber
\end{align}
with $M$ the matrix to decompose, $||\cdot||_*$ the nuclear norm defined by the sum of all singular values as a surrogate for low-rank \citep{Yuan2009}, $||\cdot||_1$ the $l_1$-norm defined by the element-wise sum of all absolute values as a surrogate for sparsity \citep{Yuan2009} and $\lambda$ a trade-off between the sparse and low-rank components to be recovered. Recent advances have shown that both components $L$ and $S$ can be recovered exactly from $M$  \citep{Candes2011,Yuan2009,Chandrasekaran2011}. In addition, recoverability is independent of the magnitude of outliers, it rather depends on the sparsity of the outliers \citep{Lin2016}.\par

The problem in (\ref{LpSeq}) can be solved computationally efficient with the alternating directions method (ADM, \citep{Yuan2009}), a method based on augmented Lagrange Multipliers \citep{Lin2010}. One drawback of this algorithm is the reliance on the nuclear norm $||\cdot||_*$ as an approximation of matrix rank. The nuclear norm is essentially an $l_1$-norm of the singular values of the matrix which over-penalizes the larger singular values \citep{Kang2015b}. This biased estimation is avoided by using a non-convex rank $||\cdot||_\gamma$ which is a closer approximation of the true matrix rank \citep{Kang2015b}. Here, we use the Robust PCA via Nonconvex Rank Approximation algorithm (noncvxRPCA, \citep{Kang2015b}) which is also based on the augmented Lagrange Multipliers method but uses $||\cdot||_\gamma$ rather than $||\cdot||_*$.

\begin{figure*}[tbh]
    \begin{center}
    \includegraphics[width=0.8\textwidth,trim=0 0 0 0, clip]{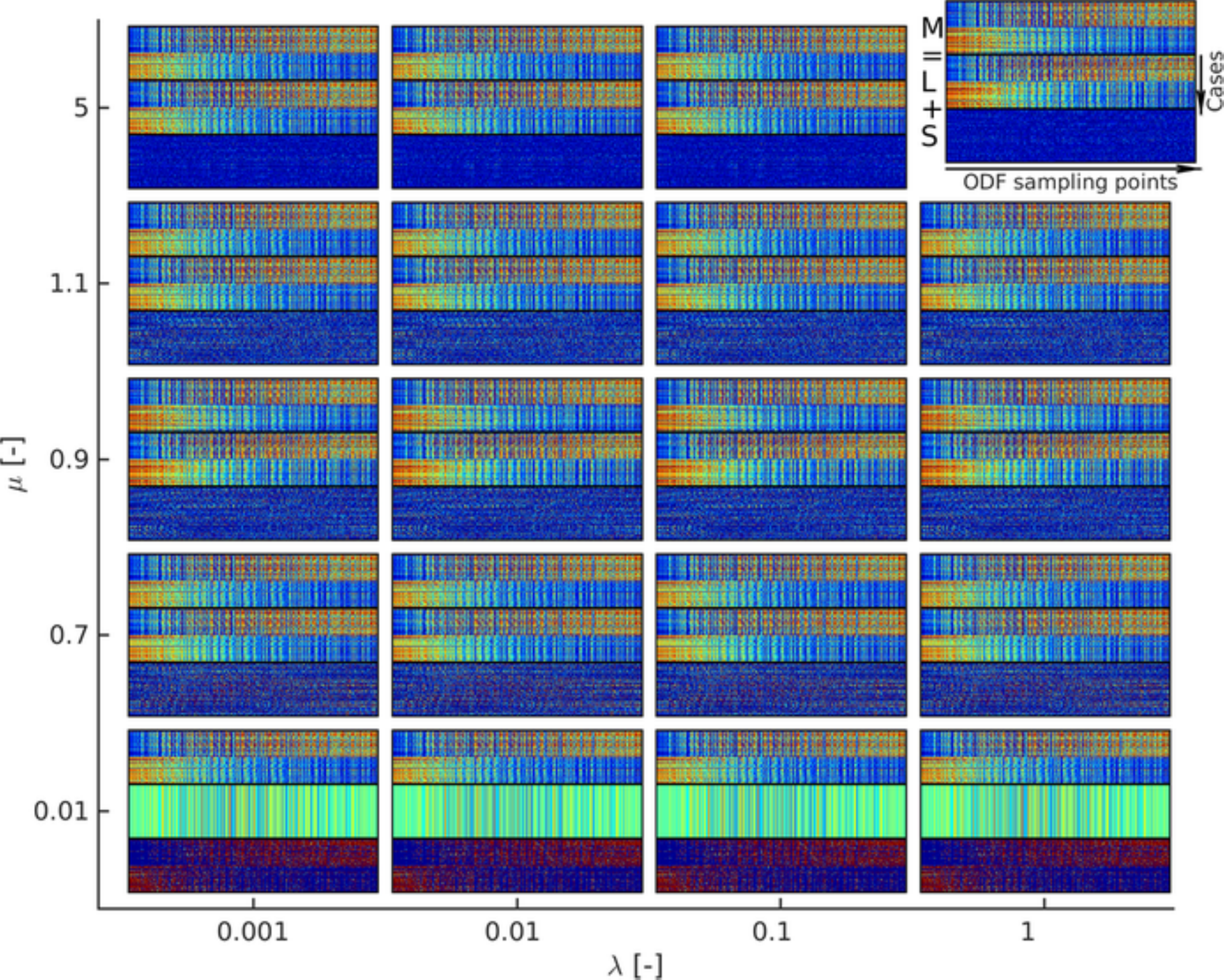}
        \caption{Impact of the choice of the regularization parameter $\lambda$ (X-axis) and the search algorithm parameter $\mu$ (Y-axis) on the performance of the $L+S$-decomposition of $M$. The performance is stable for a broad range of $\lambda$ and $\mu$. Each block of three matrices shows the ODF-matrix $M$ (top), the low-rank-matrix $L$ (middle) and the sparse matrix $S$ (bottom) (ODFs from 355 healthy volunteers (HCP)). \label{LpSparamsearch}}
    \end{center}
\end{figure*}

The $L+S$-matrix decomposition solved using the noncvxRPCA-algorithm has two main tunable parameters: $\lambda$ (Eq. \ref{LpSeq}) and $\mu$, a Lagrange penalty parameter. The parameter $\lambda$ balances $L$ and $S$ in (\ref{LpSeq}), a higher $\lambda$ will put more emphasis on the sparsity of $S$ while a lower $\lambda$ will force the rank of $L$ down. Although the outcome of (\ref{LpSeq}) could be expected to depend on the choice of $\lambda$, it was shown mathematically that a whole range of $\lambda$ values ensure the exact recovery of $L$ and $S$ \citep{Candes2011}. A universal choice of $\lambda = 1/\sqrt{n}$ with $n = max(n_1,n_2)$ and $n_1$, $n_2$ the dimensions of $M$ has been suggested \citep{Candes2011,Lin2010} and successfully applied in a large number of applications. We have employed $\lambda = 1/\sqrt{n} \approx 0.05$ in this work when observing 321 vertices per ODF and $\pm$350 subjects, though a wide range of $\lambda$ performs as desired (Fig. \ref{LpSparamsearch}).\par

 The variable $\mu$ in the augmented Lagrange multiplier optimization approach is the penalty parameter for the violation of the linear constraint $||L+S-M||$ during the search. Simulations (Fig. \ref{LpSparamsearch}) suggest that a large $\mu$ enforces a very sparse $S$ whilst a small $\mu$ decreases the rank of $L$. Hence it is important to select an appropriate value of $\mu$ for our application. Here, we have chosen to use $\mu = 0.9$ in accordance with literature \citep{Kang2015b}. Indeed, simulations (Fig. \ref{LpSparamsearch}) show that values of $\mu$ in a range around 1 balance the sparsity of $S$ with the rank of $L$ as desired.\par

Mathematical analysis has shown that the $L$ and $S$ components of $M$ can be recovered exactly with a high probability when the rank of $L$ is low and $S$ is sparse \citep{Candes2011,Yuan2009,Chandrasekaran2011,Zhou2014}. The limit for the average normalized rank of $L$ was identified by Candes, et al. \citep{Candes2011} as $rank(L)/min(n_1,n_2)\,$ $\,\leq c_1 /log(n)^2\,$ $\,\approx c_1\,0.03$ with $c_1$ a positive constant. Similarly, the upper limit for the normalized cardinality, counting the non-zero elements of a matrix as a measure for sparsity, is $m(S)/(n_1 n_2)\,$ $\,\leq c_2$ with $c_2$ a positive constant \citep{Candes2011}. While the constants $c_1$ and $c_2$ are not known, simulation results \citep{Yuan2009,Candes2011} indicate that the recovery of $L$ and $S$ is valid for normalized rank values below $0.1$ and normalized cardinalities below $0.2$. The normalized rank of $L$ and cardinality of $S$ averaged over the whole brain in this work are $0.016\,\pm\,0.002$ and $0.01\,\pm\,0.01$ respectively. Additionally, the oversampling ratio \citep{Zhou2014} of the single voxel ODF matrices M is $24\,\pm\,5$. Hence, the $L+S$-decomposition can be reliably used to identify the low-rank subspace of ODFs in a matrix of vectorized ODFs.

Once the low-rank subspace of ODFs is identified in each voxel, we use the PC-scores in these low-rank bases as input for statistical testing (Fig. \ref{LpSScheme}). The large numbers of multiple comparisons are corrected using the Threshold-Free Cluster Enhancement method (TFCE, \citep{Smith2009}) which obviates the need for a suitable cluster threshold choice \citep{Nichols2001}. In addition, combining TFCE permutation inference with complex General Linear Models (GLM) allows accounting for nuisance variables \citep{Winkler2014}. These methods assume that the joint probability distribution of the variables does not change if they are rearranged \citep{Winkler2014}. This is a valid assumption since the joint distribution of errors of the ODFs PC-scores can be assumed to be invariant on exchange \citep{Winkler2014}. Significance of group differences or correlations can be assessed based on the p-values fully corrected for multiple comparisons across space \citep{Smith2009} calculated from the TFCE-output.\par

In addition to group difference significance, the low-rank basis of the ODFs in each voxel is used to calculate difference ODFs $\Delta_{ODF}$ between subject groups $A$ and $B$ ($n_A, n_B$ members) based on the Principal Components ($PC_i$) and their PC-scores ($t_{i,j}$)
\begin{equation}
\Delta_{ODF} = \sum\limits_{i,p_i < p_{thres}} PC_i \left( \frac{1}{n_A}\sum_{j \in A}{t_{i,j}} - \frac{1}{n_B}\sum_{j \in B}{t_{i,j}}\right)\label{diffODF}
\end{equation}
for the PCs which were detected to hold significant differences $p_i < p_{thres}$ between groups. Similarly, when observing trends related to a demographic or behavioral variable, the correlation ODF $R_{ODF}$ can be calculated as
\begin{equation}
R_{ODF} = \sum\limits_{i,p_i < p_{thres}} PC_i r_i\label{diffODFtrend}
\end{equation}
with $r_i$ the correlation coefficient between $t_{i,j}$ and the demographic or behavioral variable. The ODFs in (\ref{diffODF}) and (\ref{diffODFtrend}), obtained from statistical analysis, are expressed in the same physical quantities as the original ODFs since they are expressed in terms of the PC-basis. They can be used for visualization of the significant differences between subject groups, significant trends in the dataset or as a basis for tractography visualization (Fig. \ref{LpSScheme}).\par

$\Delta_{ODF}$ and $R_{ODF}$ illustrate group differences or correlations, both in magnitude and direction, in the underlying diffusion properties of the fiber bundle in the voxel. Since the spatial extent of each peak is related to the Quantitative Anisotropy (QA \citep{Yeh2010}), both increases in $D_{ax}$ and decreases in $D_{rad}$ will increase the peak length. Similarly, both decreases in $D_{ax}$ and increases in $D_{rad}$ will decrease the peak length. Hence, in the difference ODFs, we encapsulate both possible changes, possibly missing compensating changes.

\section{Methods}

\subsection{Simulated ODF generation}
Single voxel groups of Radial DSI datasets of two crossing fiber bundles with equal weight (60°, $\lambda_1$/$\lambda_2$/$\lambda_3$ 1.00/0.10/0.10$\mu$ m$^2$/ms) and a water pool (10\%) are simulated with Radial q-space sampling (59 radial lines, 4 shells, $b_{max}$ = 4000$\,$s/mm$^2$, \citep{Baete2015RDSI}). Rician noise (SNR 30 in non diffusion attenuated signal) and group-outliers (10\%, SNR 5\%) are added to the simulated diffusion signals before reconstructing the ODFs \citep{Tuch2004}. Each single voxel group contains 100 ODFs, simulating a study with 100 co-registered cases per group. The group differences are emulated by changing $D_{ax}$ ($\lambda_1$) or $D_{rad}$ (($\lambda_2$ + $\lambda_3$)/2) of one of the two fibers fiber or the crossing angle of the fibers. Since these are single-voxel simulations, two-sided Student's t-test (5$\%$ significance level) statistics and p-values are used to evaluate the detectability of simulated group differences. The average ODFs of each group are plotted where appropriate.\par

\subsection{In vivo acquisitions}

\textit{In vivo} subject datasets were downloaded from the Human Connectome Project (HCP) consortium led by Washington University, University of Minnesota, and Oxford University. We used the 355 subjects from the December 2015 release (S900, 180/175 female/male, $28.2\,\pm\,3.9$ y/o, BMI $26.6\,\pm\,5.2$). Diffusion imaging using mono-polar gradient pulse sequence (6 b$_0$-images and 270 q-space samples on three shells, $b\,=\,$1000,2000 and 3000$\,$s/mm$^2$; all diffusion directions are acquired twice, one with phase encoding left-to-right and once with phase encoding right-to-left; TR/TE = 5500/89.50ms, 1.25 mm isotropic resolution, 210$\times$180 mm field of view, 111 slices, simultaneous multi-slice acceleration of 3 \citep{Sotiropoulos2013}; acquisition time of approximately 55 min) and structural imaging (MPRAGE; TR/TE = 2400/2.14ms, 192 slices, 1$\times$1$\times$1 mm resolution, TI = 900/1000ms, parallel imaging (2x, GRAPPA), 5:03min) was performed on a Siemens 3T Skyra with 100 mT/m maximum gradient strength.\par

Preprocessing of the HCP dataset was performed by the Human Connectome Project consortium as described in \citep{Glasser2013}. The diffusion datasets were reconstructed with the generalized q-space diffeomorphic reconstruction \citep{Yeh2010,Yeh2011} as implemented in DSIStudio \citep{Yeh2010} to the MNI-atlas \citep{Alexander2001,Lepore2010,Yeh2011,Raffelt2011,Raffelt2012}. The ODFs calculated from these reconstructions are transformed Spin Density Functions (SDF, \citep{Yeh2010,Yeh2011}) as they are multiplied by the spin density as estimated from the b0-map \citep{Yeh2011}. The spin density of the ODFs is then scaled relative to the amount of diffusing spins in 1 mm $^3$ free water diffusion, estimated as the diffusion in the cerebrospinal fluid \citep{Yeh2010}. This allows for a unified reference within and between subjects \citep{Yeh2010}.\par

\begin{figure*}[tbh]
    \begin{center}
    \includegraphics[width=0.80\textwidth,trim=0 0 0 0, clip]{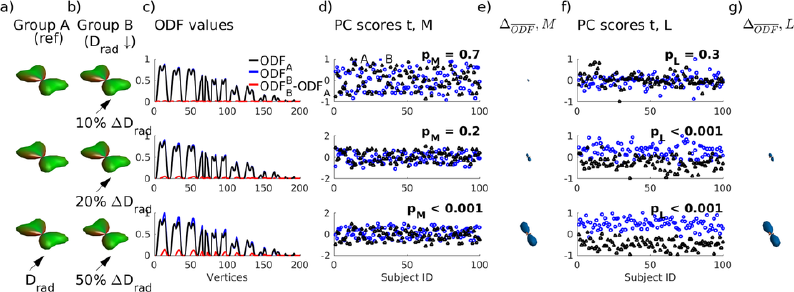}
        \caption{Simulation of detection of reductions in Radial Diffusivity ($D_{rad}\downarrow$ leads to QA$\uparrow$) of one fiber in a pair of crossing fibers (a,b). Large reductions of $D_{rad}$ (50\%) in group B (b) relative to group A (a) are easily detected by direct Principal Component Analysis (PCA) of the ODFs (d), though smaller differences (e.g. 10\% drop) are better identified by PCA of the Low-Rank $L$-matrix (f). The difference ODFs $\Delta_{ODF}$ (e,g) identify the increase of group B relatve to group A.\label{demosim}}
    \end{center}
\end{figure*}

\subsection{Simulations using in vivo acquisitions}

The choice of parameters and the ODF group difference detection of the $L+S$-matrix decomposition is evaluated using registered \textit{in vivo} ODFs selected from the HCP datasets. The ODFs from a single voxel assist in evaluating the impact of the tunable parameters $\lambda$ and $\mu$ of the noncvxRPCA algorithm. In a larger selected region, ODFs from adjacent voxels are reorganized to simulate group differences. Group differences detected with the $L+S$-\-decom\-position are then compared to the Jensen-Shannon Divergence (JSD) \citep{Cohen-Adad2011} between the mean ODFs of the voxels.\par

\subsection{Statistical testing}

Statistical significance of detected group differences is assessed using a 2-sample t-test for single voxel comparisons and simulations. In whole brain analysis, the FWE are corrected using the TFCE-method \citep{Smith2009} using the \textit{randomise} implementation of FSL. In this permutation-based testing, the nuisance variables age and gender are accounted for by using a General Linear Model (10000 permutations, \citep{Winkler2014}). Results are displayed using Matlab (Mathworks) and DSIStudio \citep{Yeh2010}.\par

Whole brain ODFs are correlated with a number of demographic and neurocognitive variables from the HCP Data Dictionary\footnote{\url{https://wiki.humanconnectome.org/display/PublicData/HCP+Data+Dictionary+Public-+500+Subject+Release}}. The analysis focuses on Body Mass Index (BMI), motor functioning (NIH Toolbox 2-minute Walk Endurance Test \textit{Endurance\_AgeAdj} and 4-meter Walk Gait Speed Test \textit{GaitSpeed\_Comp} \citep{ATS2003,Reuben2013}), language and vocabulary comprehension and decoding (NIH Toolbox Picture Vocabulary Test \textit{PicVocab\_AgeAdj} and Oral Reading Recognition Test \textit{ReadEng\_AgeAdj} \citep{Gershon2014}), episodic memory (NIH Toolbox Picture Sequence Memory Test \textit{PicSeq\_AgeAdj} \citep{Dikmen2014}) and fluid intelligence (Raven's Progressive Matrices: Number of Correct Responses \textit{PMAT24\_A\_CR} and Total Skipped Items \textit{PMAT24\_A\_SI} \citep{Bilker2012}).\par

\begin{figure}[tb]
    \begin{center}
    \includegraphics[width=0.4\textwidth,trim=0 0 0 0, clip]{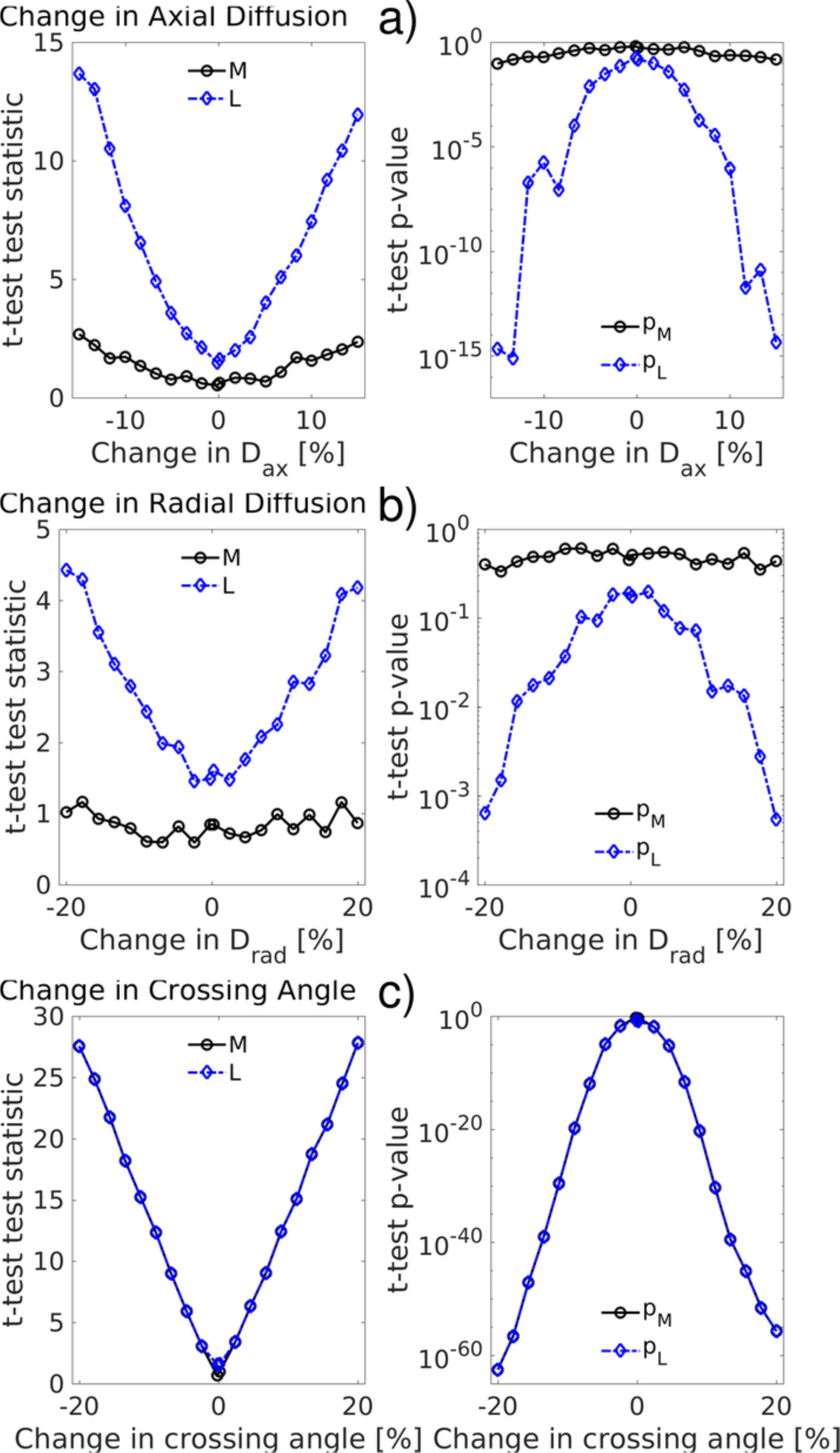}
        \caption{Simulations of percentual changes in Axial Diffusion $D_{ax}$ (a), Radial Diffusion $D_{rad}$ (b) and crossing angle (c) of ODFs of two crossing fibers. Two-sided t-test test statistics (left column) and p-value (right column) of the Principal Component-scores of the $M$ and $L$-matrices are plotted relative to percentual changes in fiber characteristics.\label{LpSsim}}
    \end{center}
\end{figure}

\begin{figure*}[tbh]
    \begin{center}
    \includegraphics[width=0.7\textwidth,trim=0 0 0 0, clip]{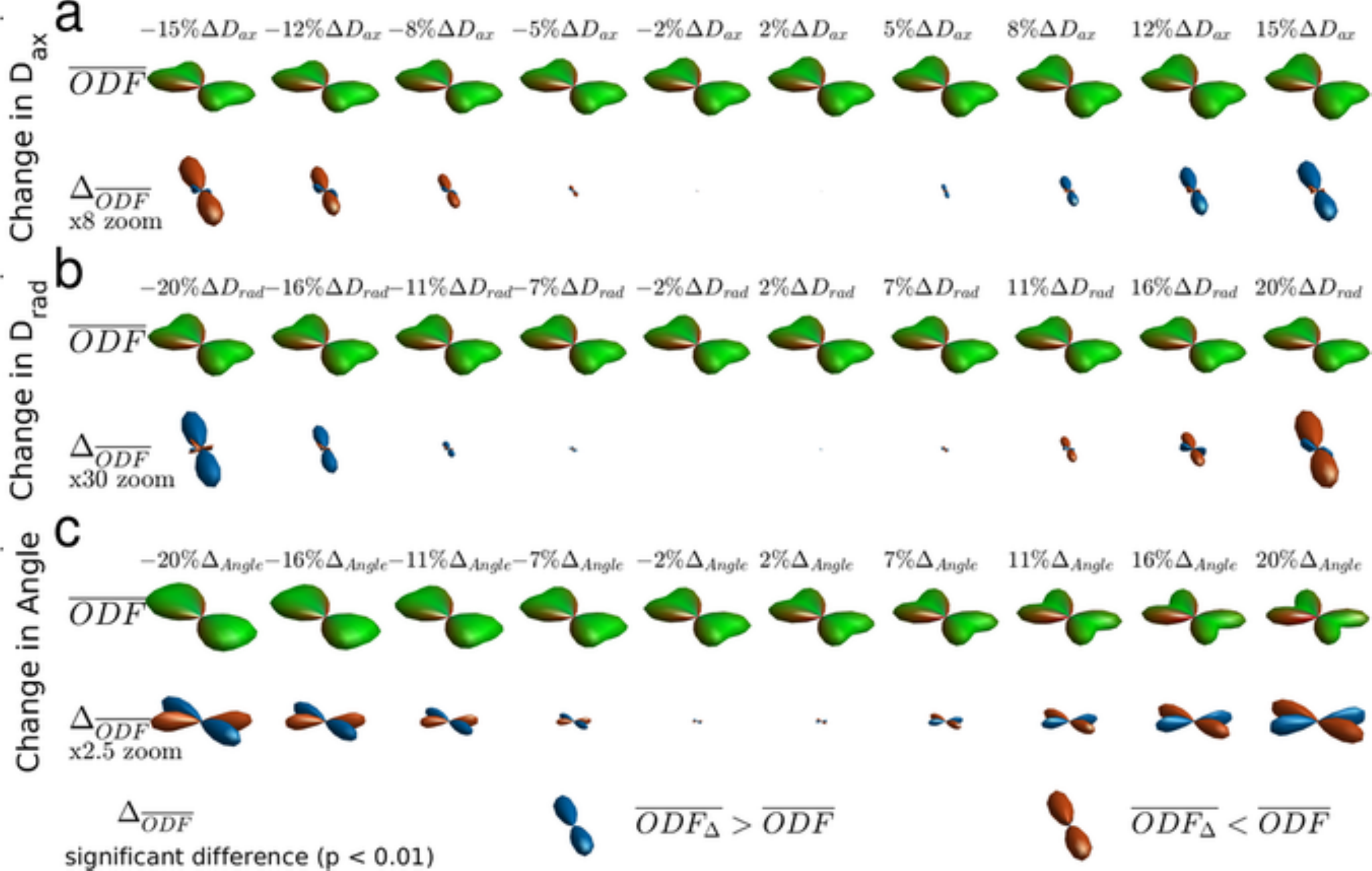}
        \caption{Difference ODFs $\Delta_{ODF}$ of simulations of two groups of two crossing fibers. A group of ODFs (average $\overline{ODF}$ displayed, green) undergoes percentual changes in Axial Diffusion $D_{ax}$ (a), Radial Diffusion $D_{rad}$ (b) and crossing angle (c) of fibers relative to the reference group of ODFs. Blue and red $\Delta_{ODF}$-lobes indicate positive and negative changes respectively.\label{diffODFsim}}
    \end{center}
\end{figure*}

\begin{figure*}[!ht]
    \begin{center}
    \includegraphics[width=0.9\textwidth,trim=0 0 0 0, clip]{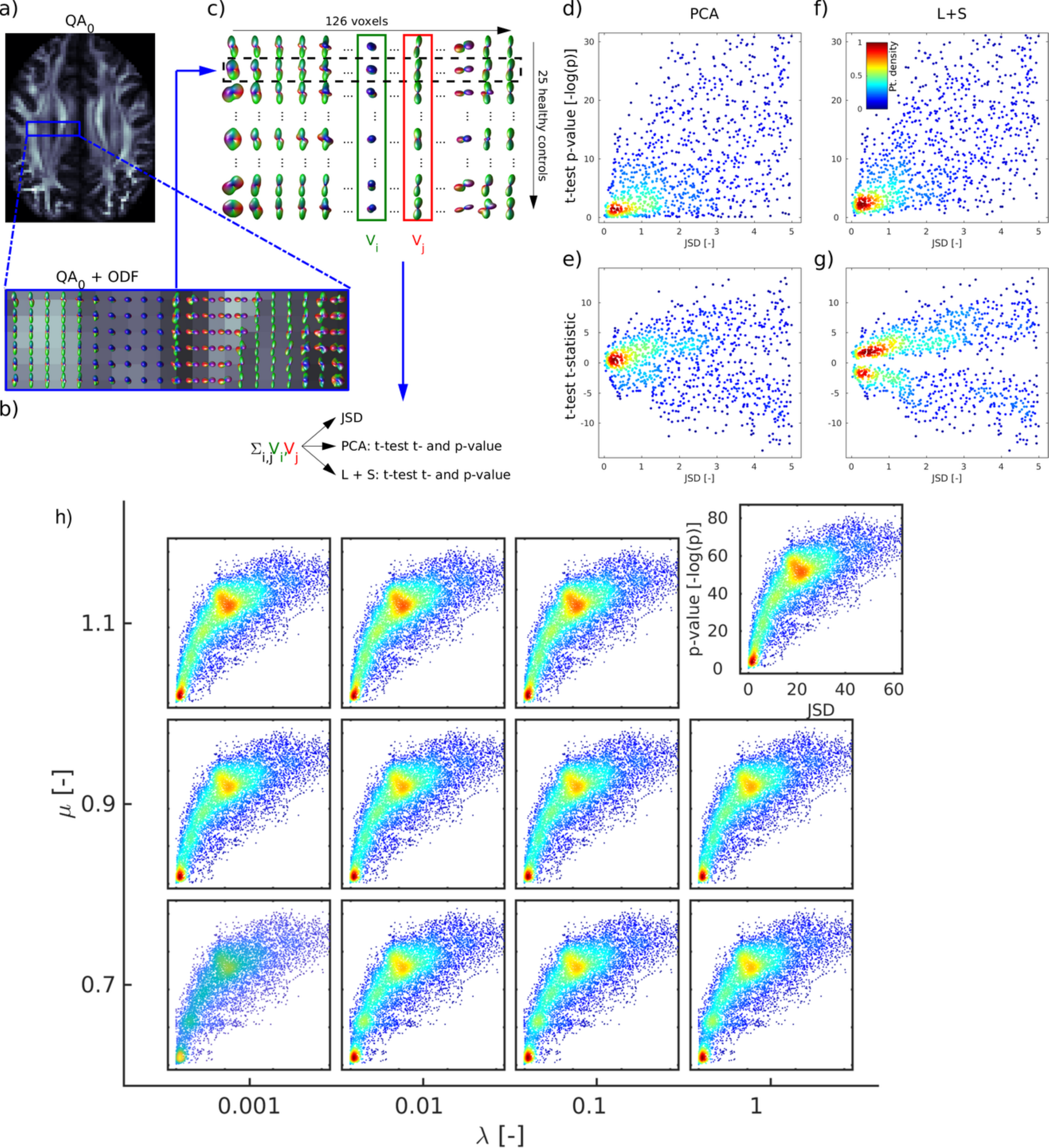}
        \caption{Evaluation of ODF group difference detection on groups of registered in vivo ODFs (c) taken from a segment (b) of a whole brain scan (a) (ODFs from 25 healthy volunteers (HCP)). The ODFs are reorganized (c) and for all possible voxel combinations (columns in the matrix in (c)), the JSD (JSD $<$ 5) is plotted versus the t-test p-value (d) and t-statistic (e) of $M$ (PCA-analysis) and the p-value (f) and t-statistic (g) of $L$ ($L+S$-analysis) (color indicates point density according to the scale in (f)). h) Impact of the choice of the parameters $\lambda$ (regularization) and $\mu$ (search algorithm penalty) of the $L+S$-decomposition on the detection of significant ODF group differences (individual plots similar to f).\label{LpSpvsJSD}}
    \end{center}
\end{figure*}

\subsection{Comparison with existing methods}

The ODFs found to correlate with BMI with ODF $L+S$ are compared with results from existing methods \citep{Smith2006,Jbabdi2010,Yeh2011,Raffelt2012,Raffelt2015}. The default TBSS pipeline \citep{Smith2006,Jbabdi2010} was used by performing registration, skeletonisation and statistical analysis as suggested by the TBSS user guide\footnote{\url{https://fsl.fmrib.ox.ac.uk/fsl/fslwiki/TBSS/UserGuide}} with FSL v5.0.9. Connectivity-based fixel enhancement (Fixel) \citep{Raffelt2012,Raffelt2015} analysis included estimation of the individual fiber response functions, calculation of individual fODFs, generating a study-specific fODF template and performing tractography on this template, registering all subjects to the template and statistical analysis of the apparent fiber density according to the MRtrix user guide\footnote{\url{http://mrtrix.readthedocs.io/en/latest/fixel_based_analysis/mt_fibre_density_cross-section.html}}. MRtrix3\footnote{\url{https://github.com/MRtrix3}} version 0.3.15 was compiled from source. Lastly, a local connectome based statistical analysis \citep{Yeh2016} was performed using DSIStudio (\citep{Yeh2010} compiled from source on March 18th, 2017). The suggested workflow\footnote{\url{http://dsi-studio.labsolver.org/Manual/diffusion-mri-connectometry}} was followed to create a connectometry database using q-space diffeomorphic reconstruction to the HCP-842 template \citep{Yeh2011} and to run group connectometry analysis (10000 permutations, T-score threshold 1.6 to obtain p $<$ 0.001).

\section{Results}

In this section, we will demonstrate the applicability of the $L+S$-matrix decomposition for detection of ODF group differences and correlation with biological and behavioral measurements using simulated and \textit{in vivo} group data.\par

\subsection{Simulation results}


Figures \ref{demosim} and \ref{LpSsim} show results of the detection of differences between groups of 100 simulated ODFs of two crossing fibers. When comparing large differences (Fig. \ref{demosim}ab, bottom row, $50\%$ reduction of $D_{rad}$), visual comparison of the average ODF-values (Fig. \ref{demosim}c) easily confirms the separation of PC-scores of both ODF-matrix $M$ (Fig. \ref{demosim}d) and Low-Rank matrix $L$ (Fig. \ref{demosim}f). When comparing smaller group differences however (Fig. \ref{demosim}, top two rows, $10\%$ and $20\%$ reduction of $D_{rad}$), statistical testing of $M$ is not significant (Fig. \ref{demosim}d) ,while testing of $L$, after separation of the individual variability in $S$, does succeed in identifying the simulated group differences (Fig. \ref{demosim}f).\par

Figure \ref{LpSsim} looks at the detectability of a broader range of changes of $D_{ax}$ (Fig. \ref{LpSsim}a), $D_{rad}$ (Fig. \ref{LpSsim}b) and crossing angle (Fig. \ref{LpSsim}c) between groups of simulated ODFs. \mrev{R3.2}{Similarly, figure \LpSsimSuppl{} studies the detectability of changes in number of fibers (Fig. \LpSsimSuppl{a,b}) and relative fiber weights (Fig. \LpSsimSuppl{c}). The left columns of Figure \ref{LpSsim} and \LpSsimSuppl{} plot} the t-test test statistic and the right columns the t-test p-value. Changes in $D_{ax}$, $D_{rad}$ \mrev{R3.2}{and relative weight}, which create more subtle differences in the ODF peaks, are better detected by analyzing the PC-scores of $L$ versus $M$. Larger changes in the ODF, such as a shift in the ODF peak orientation by changing the crossing angle \mrev{R3.2}{or adding fibers to the ODF}, are detected equally well by both approaches. Note that there are almost linear relationships between the test statistics and the changes in diffusion parameters (Fig. \ref{LpSsim}, \mrev{R3.2}{\LpSsimSuppl{}}, left column).\par

The $\Delta_{ODF}$ (Eq. \ref{diffODF}) in Fig. \ref{diffODFsim} are a visual representation of the detected ODF group differences. They scale with the detected ODF group difference for simulated changes in $D_{ax}$ (\ref{diffODFsim}a) and $D_{rad}$ (\ref{diffODFsim}b) of one fiber bundle in a crossing pair and with changes of their crossing angle (\ref{diffODFsim}c). Both an increase in $D_{ax}$ and a decrease in $D_{rad}$ give rise to a larger peak, hence a positive (blue-colored) $\Delta_{ODF}$-lobe. A change in crossing angle gives rise to a positive (blue-colored) and a negative (red-colored) lobe scaling with the change in crossing angle (\ref{diffODFsim}c).\par

\subsection{Simulations using \textit{in vivo} data}

The relationship between \textit{in vivo} group differences and t-test statistics and p-values is explored in Figure \ref{LpSpvsJSD}.  A central segment is randomly taken from the registered \textit{in vivo} whole brain HCP dataset (Fig. \ref{LpSpvsJSD}a,b). Group differences are artificially introduced by comparing all ODFs of each pair of voxels in the segment (Fig. \ref{LpSpvsJSD}bc). The results of these comparisons using $M$ (Fig. \ref{LpSpvsJSD}de) and $L$ (Fig. \ref{LpSpvsJSD}fg) are plotted versus the JSD of the average ODFs of the respective voxels. The detected p-values are smaller (larger $-\log(p)$ values) when analyzing $L$ over $M$, indicating higher detected significance. The comparison test is in addition inconclusive (p close to 1) on less occasions. These beneficial properties of the analysis of $L$ remain for a wide range of values for $L+S$-matrix decomposition parameters $\lambda$ and $\mu$ (Fig. \ref{LpSpvsJSD}h).\par

\subsection{\textit{In vivo} results}

\begin{figure*}[tbh]
    \begin{center}
    \includegraphics[width=0.90\textwidth,trim=0 0 0 0, clip]{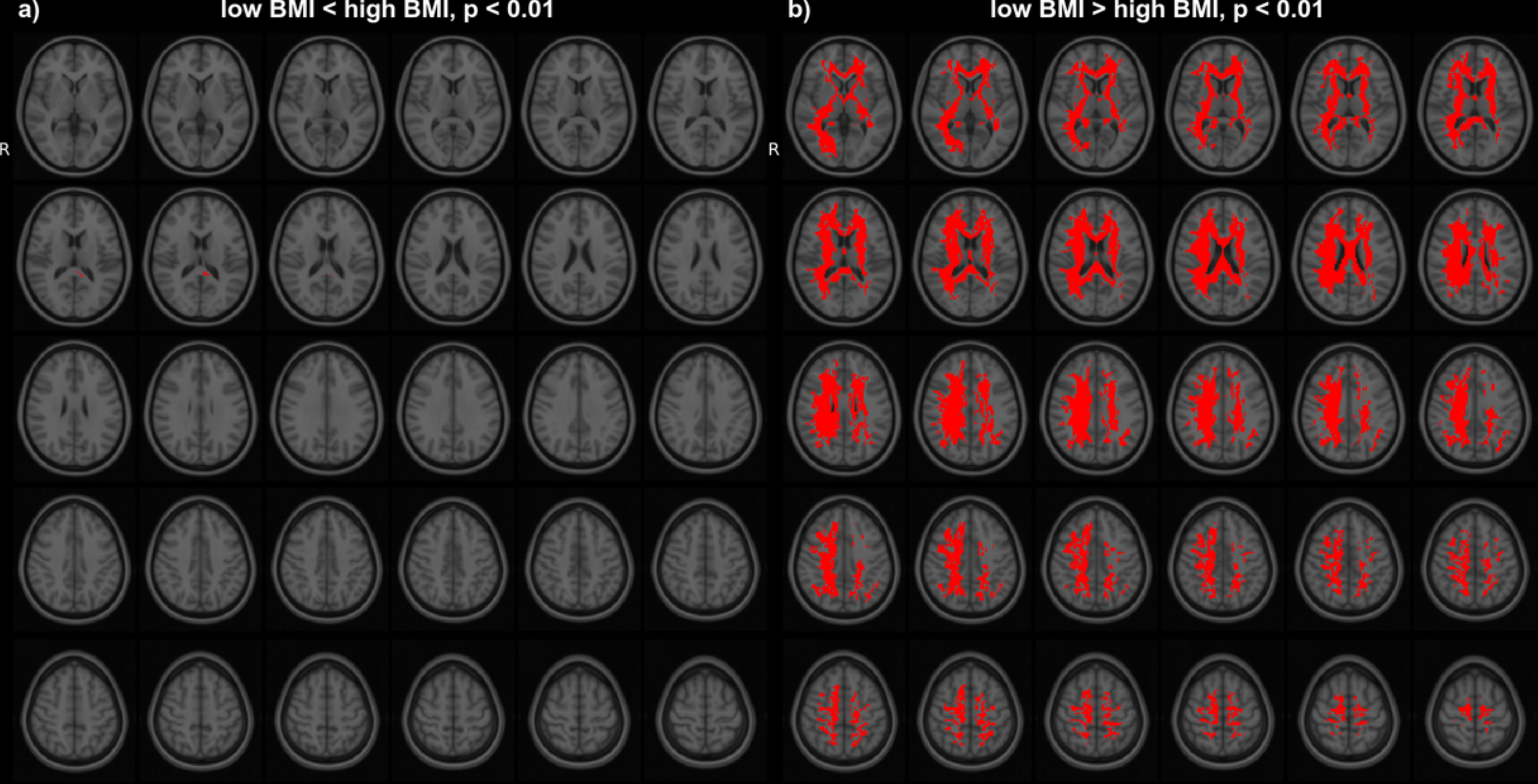}
        \caption{Voxels with ODFs significantly positively (a) or negatively (b) correlated with BMI as detected after isolating ODF-features (L) from individual variability (S) using the ODF $L+S$ approach in a cohort of healthy HCP volunteers. Voxels with FWE p-value $<$ 0.01 (red) are overlaid on the MNI-atlas.\label{noncvxRPCA_HCP}}
    \end{center}
\end{figure*}

\begin{figure*}[tbh]
    \begin{center}
    \includegraphics[width=0.9\textwidth,trim=0 0 0 0, clip]{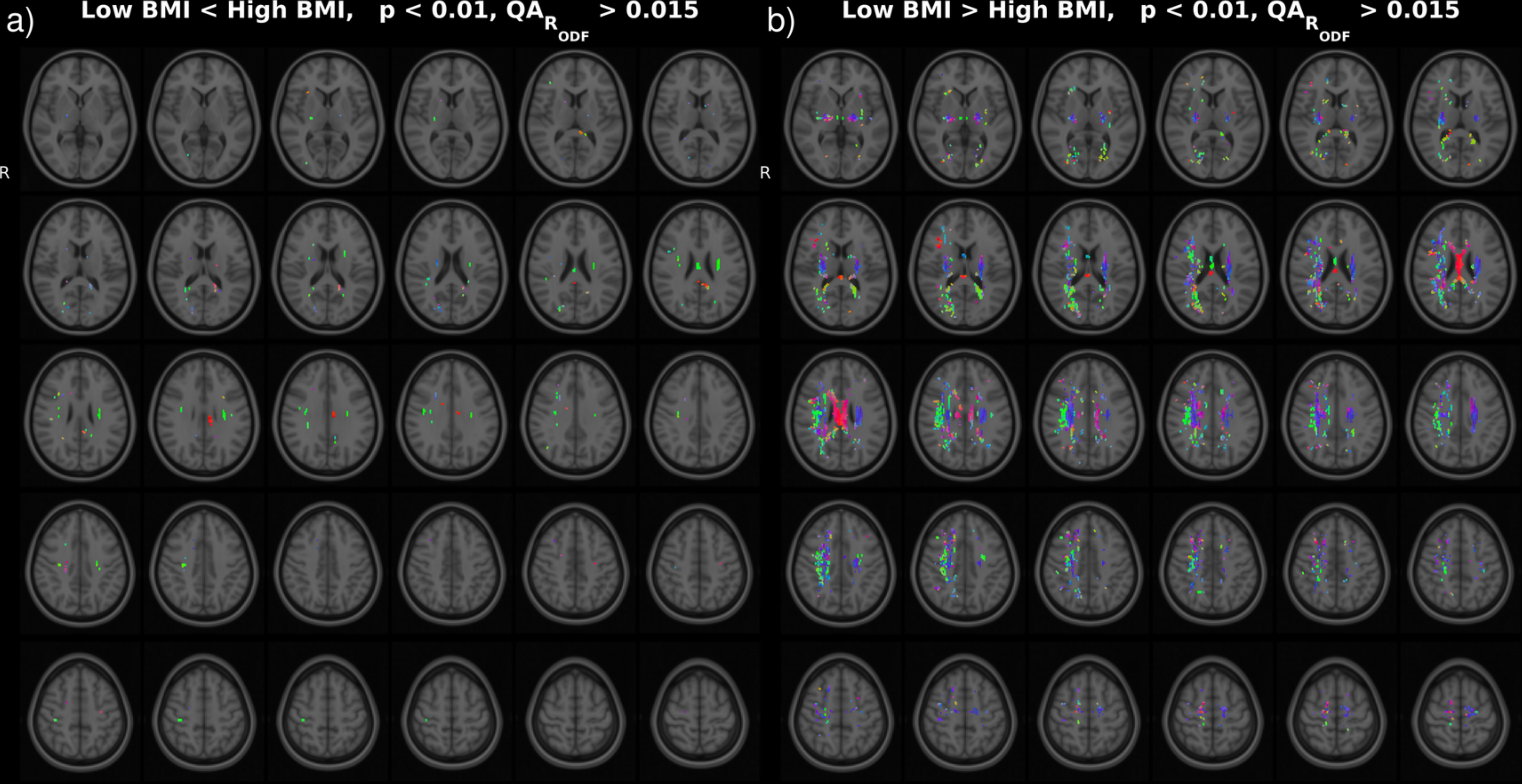}
    \includegraphics[width=0.9\textwidth,trim=0 0 0 0, clip]{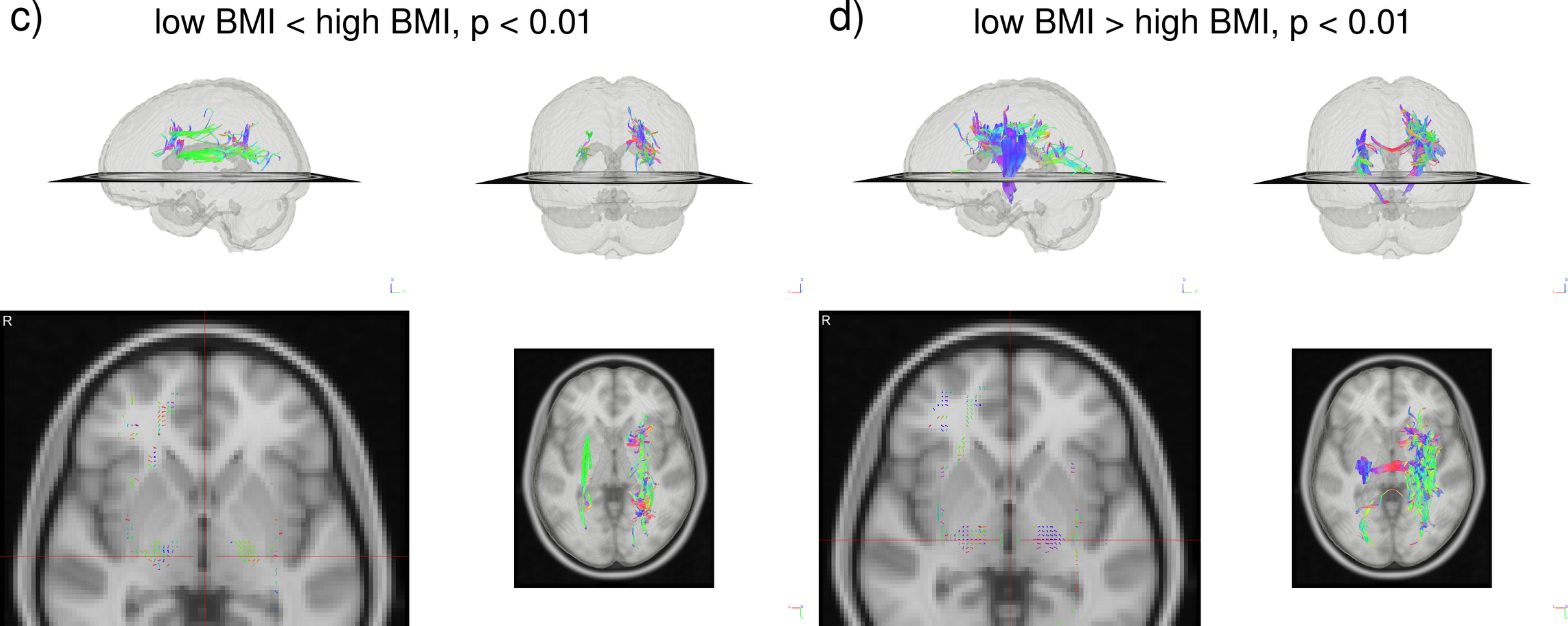}
        \caption{Fiber directions (a,b, $QA_{R_{ODF}} < 0.015$) identified from and fiber tractography of the $R_{ODF}$ (c,d, $p < 0.01$, $QA_{R_{ODF}} < 0.001$) positively (a,c) or negatively (b,d) correlating with BMI in a cohort of healthy HCP volunteers. (Eq. \ref{diffODFtrend})\label{diffODFdirHCP}}
    \end{center}
\end{figure*}

\begin{figure*}[!th]
    \begin{center}
    \includegraphics[width=0.80\textwidth,trim=0 0 0 0, clip]{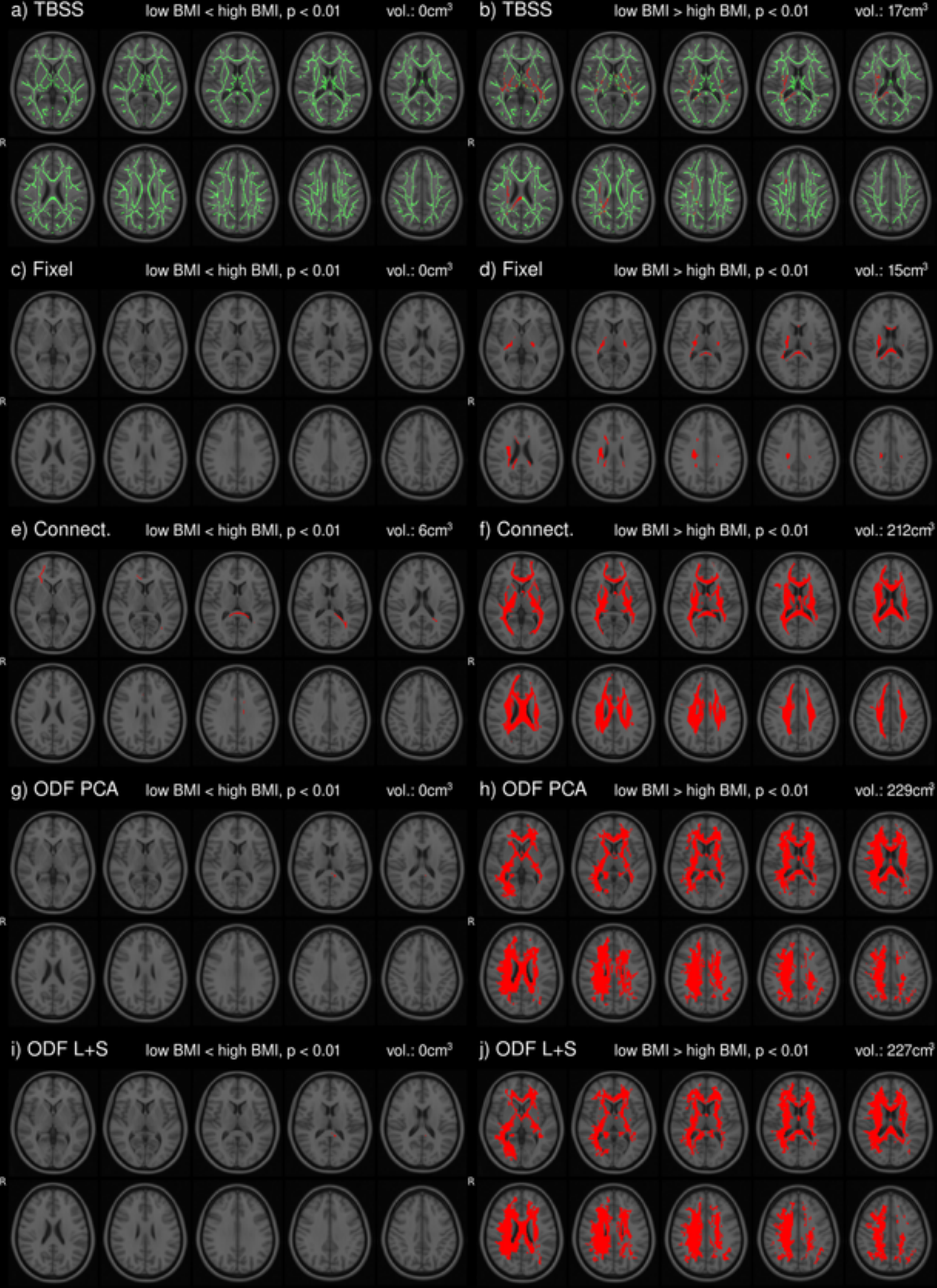}
        \caption{Voxels on selected slices significantly positively (a,c,e,g,i) or negatively (b,d,f,h,j) correlated with BMI as detected with Tract-based Spatial Statistics (TBSS, a,b), with Connectivity based fixel enhancement (Fixel, c,d), with the Connectometry based approach as implemented in DSIStudio (Connect., e,f) and with the ODF PCA and ODF $L+S$ approach in a cohort of healthy HCP volunteers. Voxels with FWE p-value $<$ 0.01 (red) are overlaid on the MNI-atlas and the mean FA skeleton (green, TBSS). The volume of positive findings is indicated at the top right of each plot. Full brain results are displayed in Fig. \ref{noncvxRPCA_HCP} (ODF $L+S$) and Suppl. Fig. \OtherMethodsBMIa{} and \OtherMethodsBMIb{b}.\label{OtherMethods_BMI_sel}}
    \end{center}
\end{figure*}

In the HCP dataset, ODFs correlate strongly with the demographic variable BMI. As expected, the ODFs negatively correlate (Fig. \ref{noncvxRPCA_HCP}b) with BMI, indicating a loss of anisotropy with increasing BMI (Fig. \ref{diffODFsim}a,b). This result is consistent with the well-established negative association between global white matter integrity and physical obesity \citep{Mueller2011, Stanek2011, Verstynen2012, Gianaros2013, Verstynen2013}. The fiber directions (Fig. \ref{diffODFdirHCP}a,b) identified from $R_{ODF}$ can be used to perform tractography (Fig. \ref{diffODFdirHCP}c,d). Resulting tracts (Fig. \ref{diffODFdirHCP}c,d) show a pronounced loss of anisotropy in the corticospinal tracts, the optic radiations and the right superior longitudinal fasciculus. These results are corroborated by the existing methods TBSS \mrev{R2.1b}{(Fig. \ref{OtherMethods_BMI_sel}a,b, \OtherMethodsBMIa{a,b})}, Connectivity-based fixel enhancement \mrev{R2.1b}{(Fig. \ref{OtherMethods_BMI_sel}c,d, \OtherMethodsBMIa{c,d})} and local connectometry \mrev{R2.1b}{(Fig. \ref{OtherMethods_BMI_sel}e,f, \OtherMethodsBMIb{a,b})}. \mrev{R2.1a}{The volume of positive findings of correlation with BMI is largest when using the full ODF information with the ODF $L+S$ approach (Fig. \ref{OtherMethods_BMI_sel}).} \mrev{R2.1c}{In addition, in a test of specificity, no voxels are found to correlate with randomly permuted BMI (Fig. \noncvxRPCAHCPBMIRand{}).} \par

\mrev{R1.2}{The individual variability of ODFs can be caused by ODF reconstruction errors due to image artifacts (Fig. \FigOutlierODF{.1}), registration errors and individual differences in brain structure (Fig. \FigOutlierODF{.3-4}). In the HCP ODFs individual variability due to the first two of these contributions is limited by the low number of artifacts and the high image resolution in the HCP DWI images.} \mrev{R1.1}{As a result, ODF correlations of BMI identified with ODF $PCA$ ((Fig. \ref{OtherMethods_BMI_sel}g,h, \OtherMethodsBMIb{c,d})) and ODF $L+S$ (Fig. \ref{OtherMethods_BMI_sel}i,j, \ref{noncvxRPCA_HCP}) are similar in this dataset. However, in a dataset with higher individual variation, notable improvement of ODF $L+S$ over ODF PCA can be observed \citep{Baete2016Whistler}.} \par

\begin{figure*}[!ht]
    \begin{center}
    \includegraphics[width=0.9\textwidth,trim=0 0 0 0, clip]{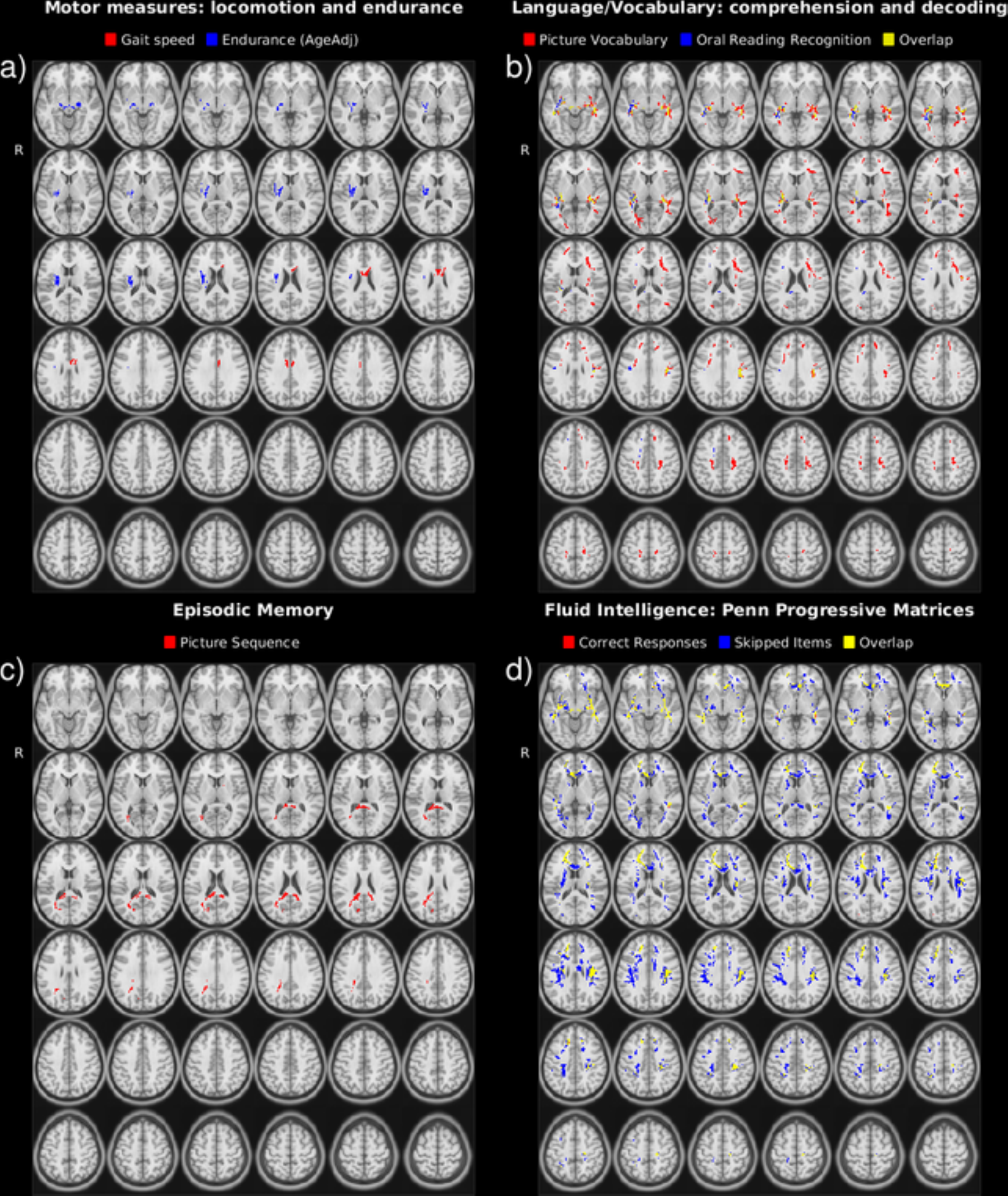}
        \caption{Voxels with ODFs significantly correlated with the indicated neurocognitive metrics in a cohort of healthy HCP volunteers as detected after isolating ODF-features (L) using the ODF $L+S$ approach. Voxels with FWE p-value $<$ 0.05 are overlaid on the MNI-atlas.\label{behavioralHCP}}
    \end{center}
\end{figure*}

\begin{figure*}[!ht]
    \begin{center}
    \includegraphics[width=0.9\textwidth,trim=0 0 0 0, clip]{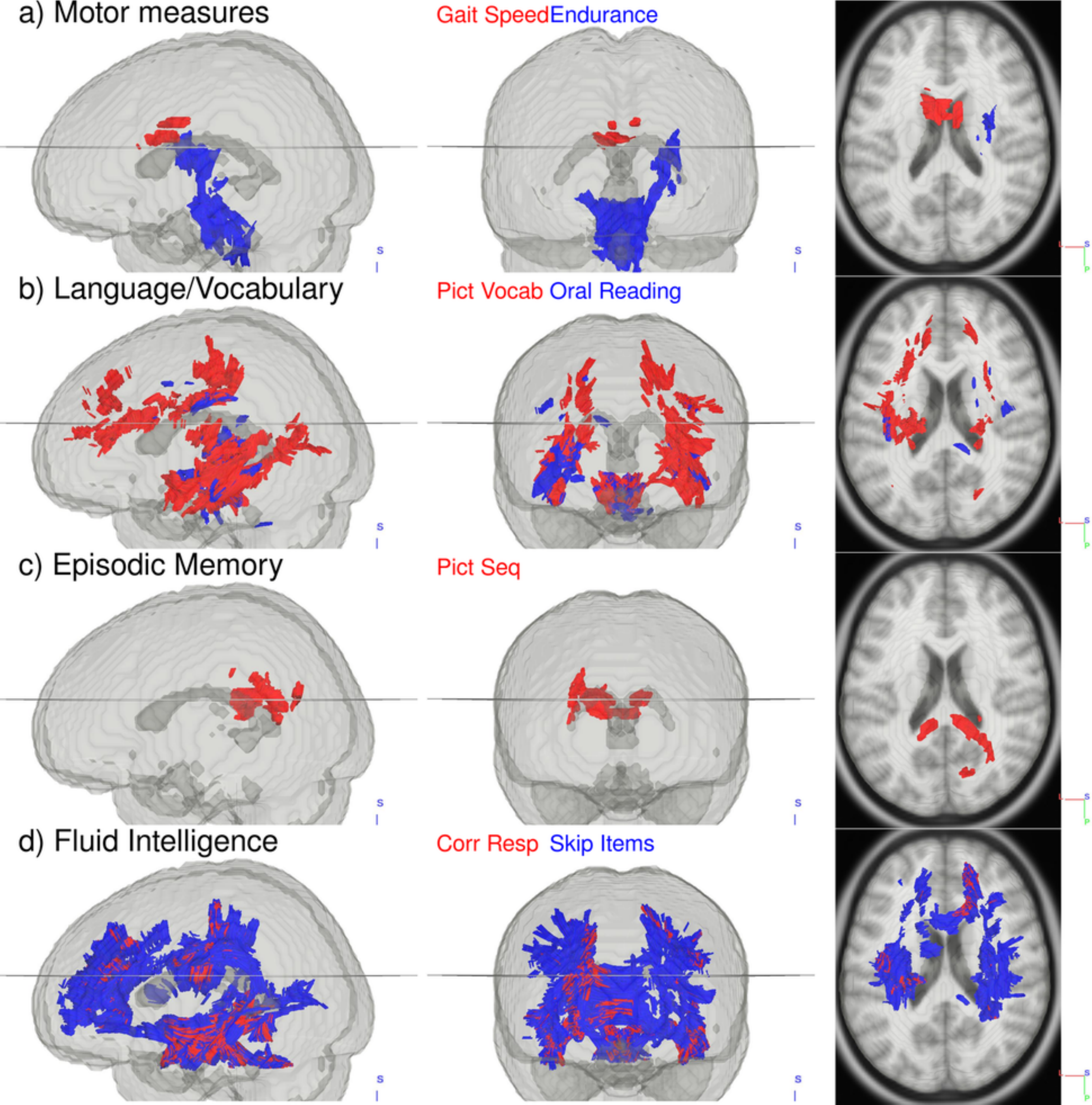}
        \caption{Fiber tracts identified by fiber tractography of the $R_{ODF}$ ($p < 0.05$, $QA_{R_{ODF}} < 0.001$) correlating with the indicated neurocognitive metrics in a cohort of healthy HCP volunteers (Eq. \ref{diffODFtrend}). Tracts are displayed in a surface and on a slice derived from the MNI-atlas.\label{behavioralTrackHCP}}
    \end{center}
\end{figure*}

The HCP ODFs also correlate with neurocognitive measures (Fig. \ref{behavioralHCP} and \ref{behavioralTrackHCP}). Walking endurance (Fig. \ref{behavioralHCP}a and \ref{behavioralTrackHCP}a) predictably relates to the corticospinal tract, while the frontal part of the corpus callosum indicated in gait speed has been identified before when comparing endurance athletes with non-athletes \citep{Raichlen2016} and when studying the effect of treadmill training after stroke \citep{Enzinger2009}. Language recognition and comprehension tasks (Fig. \ref{behavioralHCP}b and \ref{behavioralTrackHCP}b), as measured by the Oral Reading and Picture Vocabulary test, correlate with ODFs in areas identified by fMRI work \citep{Berl2010}: posterior superior temporal gyrus and inferior frontal gyrus connected by stretches of the arcuate fasciculus pathway/superior longitudinal fasciculus III. Besides these areas, language comprehension also relates with ODFs in the medial frontal gyri and the left precuneus \citep{Schmithorst2007}.
Our analysis further connects the Episodic Memory measure with the posterior cingulate and precuneus, though not with the medial temporal cortex (Fig. \ref{behavioralHCP}c and \ref{behavioralTrackHCP}c). These areas have been shown to deactivate in episodic memory tasks \citep{Dickerson2010}.
Lastly, Fluid Intelligence correlates widely with ODFs in the prefrontal, parietal and temporal cortex as indicated before \citep{Gray2003} as well as with ODFs along tracts connecting these regions (Fig. \ref{behavioralHCP}d and \ref{behavioralTrackHCP}d).

\section{Discussion}

In analyzing large neuroimaging data sets, we strive to fully utilize all the information available in each voxel to identify group differences or population correlations. The available information in each voxel of high quality diffusion MRI acquisitions is often represented as an ODF, be it a diffusion ODF or a fiber ODF. ODFs and their key features identified by PCA-analysis are thus the ideal starting point for statistical analysis. These statistical tests are however biased by individual variability of the subjects since PCA is sensitive to outliers \citep{Zhou2014,Lin2016}.  This bias is mitigated in the results presented here by separating individual variability and essential ODF features through the $L+S$ decomposition of the ODFs.\par

Statistical testing of ODFs based on $L+S$ decomposition more reliably identifies the underlying relationships between ODF shapes and observed behavior. Indeed, separating the outliers from key ODF features reduces the uncertainty regarding the existence of correlations in the dataset (Fig. \ref{LpSsim},\LpSsimSuppl{},\ref{diffODFsim},\ref{LpSpvsJSD}). This reduced uncertainty increases the detected significance of group differences (Fig. \ref{LpSsim}, \LpSsimSuppl{} and Fig. \ref{LpSpvsJSD}d,e vs. \ref{LpSpvsJSD}f,g). Furthermore, removal of individual variability decreases the bias in the estimated correlations coefficients $r$ and PC-scores $t$. We can thus deploy $r$ and $t$ to calculate difference and correlation ODFs (resp. $\Delta_{ODF}$ and $R_{ODF}$) which help in visual interpretation of the test results (Fig. \ref{diffODFsim}, \ref{diffODFdirHCP} and \ref{behavioralTrackHCP}).\par

Several methods exist to analyze populations of diffusion MRI datasets, typically working on a reduced dimensionality subset of the diffusion data. TBSS \citep{Jbabdi2010}, limiting the analysis to a projection to a tract skeleton, succeeds in identifying the tracts most significantly correlated with BMI (analysis of FA, Fig. \mrev{R2.1b}{\ref{OtherMethods_BMI_sel}a,b, \OtherMethodsBMIa{a,b}}) but misses the full extent of the correlations. The Connectivity-based fixel enhancement \citep{Raffelt2015} and Connectometry \citep{Yeh2016} approaches do include more of the available information and hence perform better than the TBSS method (Fig. \mrev{R2.1b}{\ref{OtherMethods_BMI_sel}c,d, \OtherMethodsBMIa{c,d}} (Fixel enhancement) and \mrev{R2.1b}{\ref{OtherMethods_BMI_sel}e,f, \OtherMethodsBMIb{a,b}} (Connectometry) vs \mrev{R2.1b}{\ref{OtherMethods_BMI_sel}a,b, \OtherMethodsBMIa{a,b}} (TBSS)).\par

None of the above methods however capitalizes on the full ODF information. \mrev{R2.5}{This in contrast to the approach presented in this paper. The ODF $L+S$ approach indeed identifies a larger volume of significant findings (227 cm$^3$) than the existing methods tested here (TBSS 17 cm$^3$, Connectivity-based fixel enhancement 15 cm$^3$ and Connectometry 212 cm$^3$, Fig. \ref{OtherMethods_BMI_sel}b,d,f,h,j). That is, by analyzing the full ODF information and not reducing the dimensionality of the diffusion data, as is commonly done, the ODF $L+S$ approach is able to pick up on smaller significant changes, better grasping the full extent of the significant findings. Limitations on computational power and diffusion acquisitions (e.g. DTI) which inspired the data reduction of older methods no longer exist. It is hence advisable to maximize the amount of information included in the analysis as in the approach presented here.}\par

Using the $L+S$ decomposition (Eq. \ref{LpSeq}) means that a choice for the balance parameter $\lambda$ and the Lagrange penalty parameter $\mu$ has to be made. Choosing $\lambda$ is straightforward as a universal choice $\lambda = 1/\sqrt{n}$ is suggested by theoretical considerations \citep{Candes2011,Lin2016}. Simulations furthermore show that the decomposition is relatively insensitive to $\lambda$ (Fig. \ref{LpSparamsearch}). An appropriate value for $\mu$ will however have to be found heuristically. While our results were stable for $\mu$ in range around 1, it is clear that the extraction of essential ODF features will suffer when $\mu$ is too small (too few ODF features in L) or too large (too much individual variability in L) (Fig. \ref{LpSparamsearch}).\par

Group differences or behavioral correlations of subject ODFs signal micro-structural changes in the white matter. The origin of the detected changes can unfortunately not always be teased out from the ODF analysis. For instance, both increases in $D_{ax}$ and decreases in $D_{rad}$ will increase anisotropy and as a result increase the ODF peak length. Both changes will thus results in similar $\Delta_{ODF}$ or $R_{ODF}$ (Fig. \ref{diffODFsim}), which will both illustrate the detected change in anisotropy. As a result, it is not possible to separately detect changes to axon density, axon diameter, membrane permeability and axon bundle curvature and divergence; a drawback shared with other analysis methods \citep{Raffelt2012,Yeh2016}. Group differences may consequently not be identified at all in the relatively unlikely case of confounding changes to these micro-structure parameters \citep{Raffelt2012}.\par

Besides potential confounding micro-structure changes, the presented ODF $L+S$ approach is also sensitive to \mrev{R3.5}{other limitations. The main limitations are the} processing parameters of the analysis pipeline. The choice of $L+S$ algorithm parameters $\lambda$ and $\mu$ is straightforward and the decomposition is stable in a range of parameter choices. Nevertheless, certain combinations of $\lambda$ and $\mu$ will affect the outcome of the analysis. \mrev{R3.5}{This might be avoided by future automated tuning strategies for $\lambda$ and $\mu$.} \mrev{R1.3}{In the stable parameter range, a high probability of exact recovery of $L$ from $M$ is mathematically guaranteed when $L$ is low rank and $S$ is sparse \citep{Candes2011,Yuan2009,Chandrasekaran2011,Zhou2014}. The ODF $L+S$ approach is thus limited to low rank $L$. Our \textit{in vivo} analysis shows though that $L$ is sufficiently low rank for a stable $L+S$ decomposition.} Results will further depend on the number of available diffusion directions and shells, quality control of the images, careful choice of preprocessing steps and parameters, the reconstruction used to generate the ODFs, the choice of template for registration, the method used for registration to a template and the parameters of the FWE correction method. For each of these steps, we have striven to make common sense best practice choices. \mrev{R3.3}{Of these steps, the often imperfect inter-subject registration will have the largest confounding impact, similar to other analysis methods \citep{Raffelt2012,Yeh2016}.}\par


In conclusion, high quality diffusion MRI datasets of groups of individuals open a window to studying brain structure changes related to disease condition and behavioral functions. Full incorporation of all available diffusion information however risks biasing the outcome by outliers, often leading to statistical analysis of diffusion measures with reduced dimensionality. Here we apply a Low-Rank plus Sparse decomposition on the voxelwise ODF distributions. Analyzing the Low-Rank ODF distribution reduces the impact of inter-subject variability and thus avoids outlier bias by focusing on the essential ODF features of the population while maximizing the dimensionality of included diffusion information. This approach provides a foundation for improved detection of group differences in DWI through PCA-based analysis. The identified group differences can then also be visualized with difference ODFs $\Delta_{ODF}$ and correlation ODFs $R_{ODF}$. This method will aid in the detection of smaller group differences in clinically relevant settings as well as in neuroscience applications.\par
\par
Source code (Matlab) for the ODF $L+S$ approach is available for download at \url{https://bitbucket.org/sbaete/odflpluss}.

\section*{Acknowledgements}
This project is supported in part by the National Institutes of Health (NIH, R01\-CA111996, R01\-NS082436, R01\-MH00380 and P41\-EB017183). Data were provided by the Human Connectome Project, WU-Minn Consortium (Principal Investigators: David Van Essen and Kamil Ugurbil; 1U54MH091657) funded by the 16 NIH Institutes and Centers that support the NIH Blueprint for Neuroscience Research; and by the McDonnell Center for Systems Neuroscience at Washington University.

\section*{References}

\bibliography{dsi}

\beginsupplement

  \begin{figure}[tbh]
      \begin{center}
      \includegraphics[width=0.45\textwidth,trim=0 0 0 0, clip]{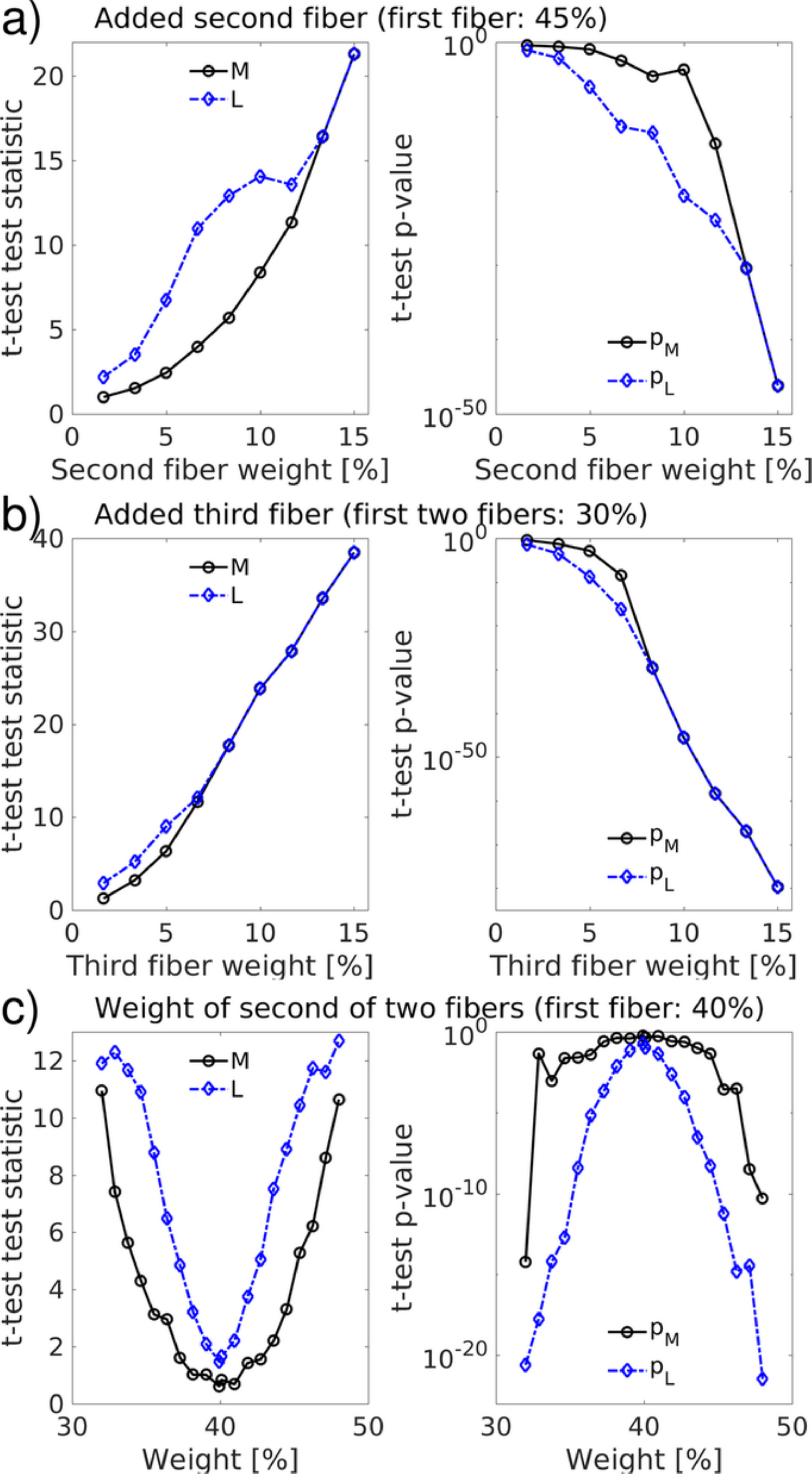}
          \caption{Simulations of changes in number of fibers (added second (a) and third (b) fiber) and relative fiber weight (c) of crossing fibers ODFs. Two-sided t-test test statistics (left column) and p-value (right column) of the Principal Component-scores of the $M$ and $L$-matrices are plotted relative to percentual changes in fiber characteristics.\label{LpSsimSuppl}}
      \end{center}
  \end{figure}

  \begin{figure*}[tbh]
      \begin{center}
      \includegraphics[width=0.80\textwidth,trim=0 0 0 0, clip]{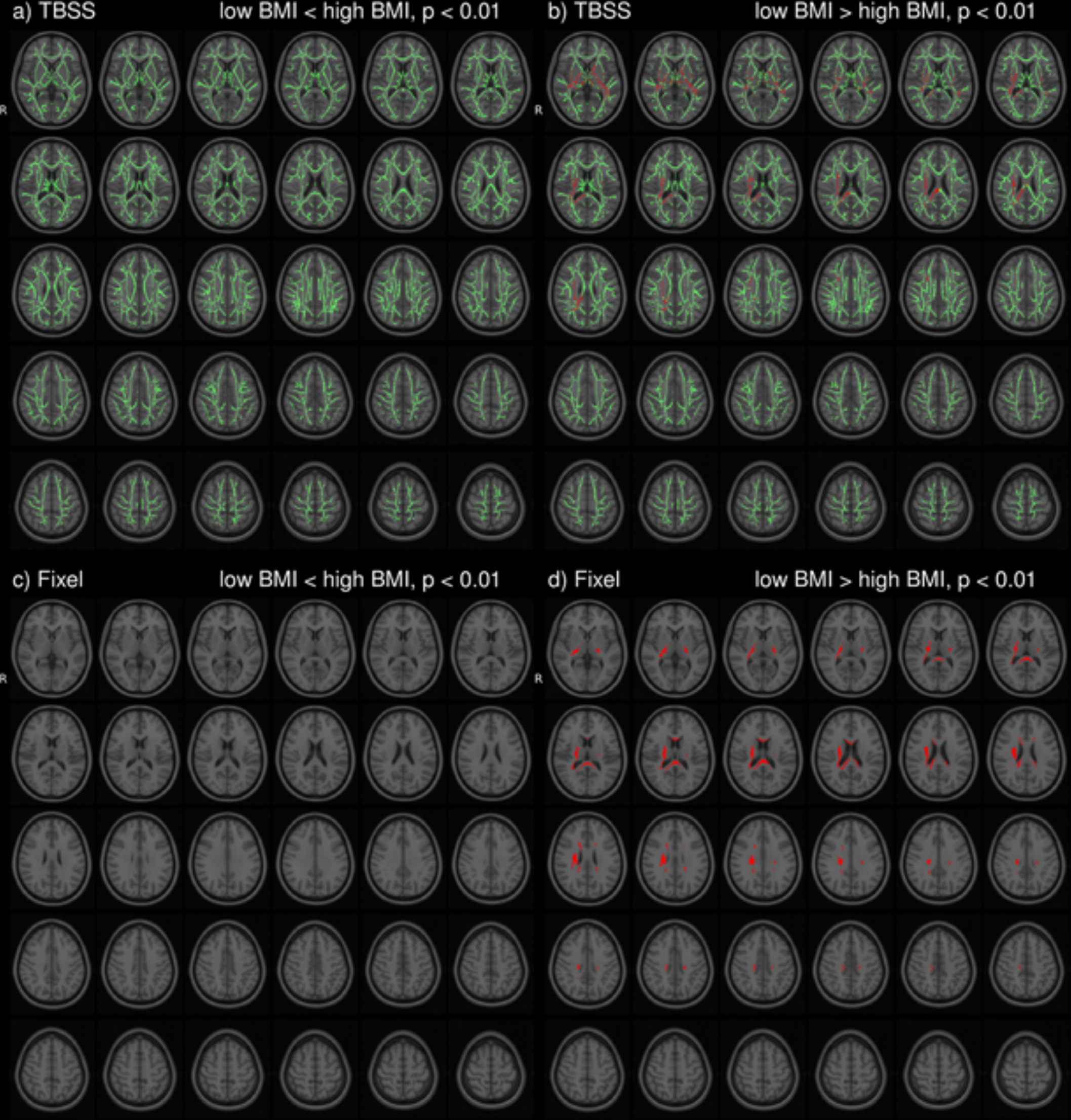}
          \caption{Voxels significantly positively (a,c) or negatively (b,d) correlated with BMI as detected with Tract-based Spatial Statistics (TBSS, a,b) and, with Connectivity based fixel enhancement (Fixel, c,d) in a cohort of healthy HCP volunteers. Voxels with FWE p-value $<$ 0.01 (red) are overlaid on the MNI-atlas and the mean FA skeleton (green, TBSS).\label{OtherMethods_BMIa}}
      \end{center}
  \end{figure*}

  \begin{figure*}[tbh]
      \begin{center}
      \includegraphics[width=0.80\textwidth,trim=0 0 0 0, clip]{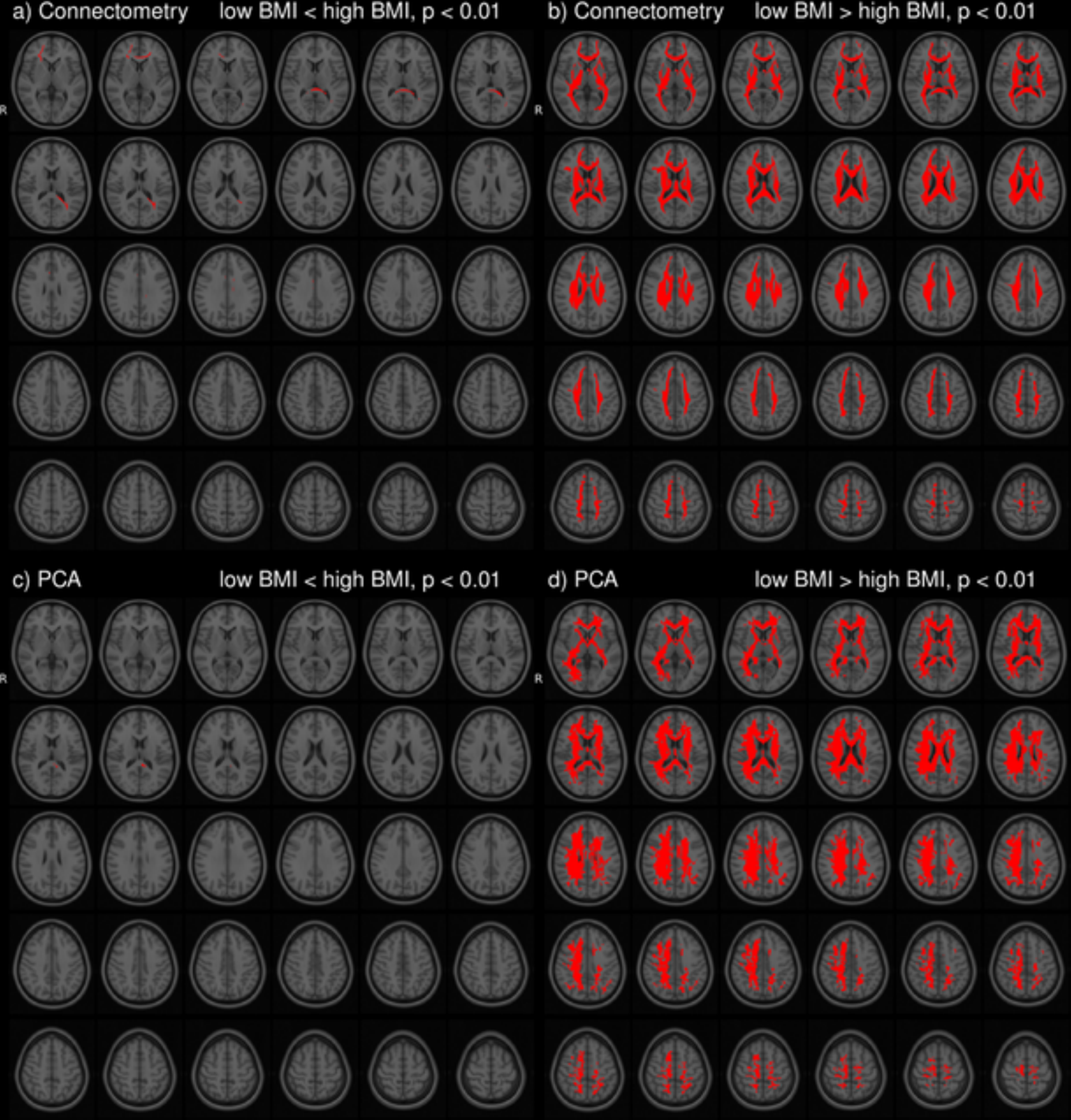}
          \caption{Voxels significantly positively (a,c) or negatively (b,d) correlated with BMI as detected with the Connectometry based approach as implemented in DSIStudio (Connectometry, a,b) and with the ODF PCA (PCA, c,d) approach in a cohort of healthy HCP volunteers. Voxels with FWE p-value $<$ 0.01 (red) are overlaid on the MNI-atlas and the mean FA skeleton (green, TBSS).\label{OtherMethods_BMIb}}
      \end{center}
  \end{figure*}

  \begin{figure*}[tbh]
      \begin{center}
      \includegraphics[width=0.80\textwidth,trim=0 0 0 0, clip]{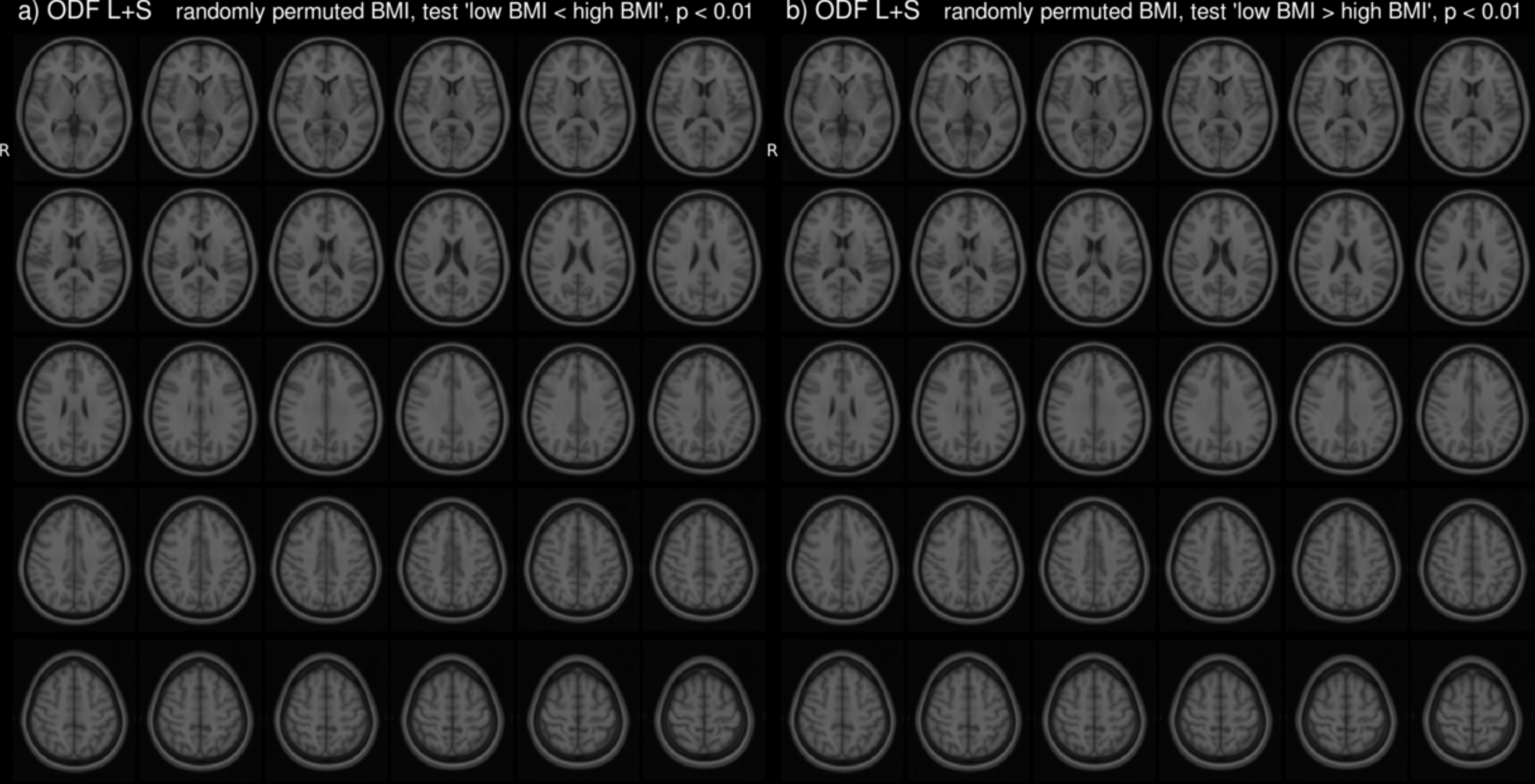}
          \caption{Analysis similar to Fig. \noncvxRPCAHCP, but with randomly permuted BMI. Voxels significantly positively (a) or negatively (b) correlated with randomly permuted BMI with FWE p-value $<$ 0.01 (red) using the ODF $L+S$ approach in a cohort of healthy HCP volunteers are overlaid on the MNI-atlas.\label{noncvxRPCA_HCPBMIRand}}
      \end{center}
  \end{figure*}

  \begin{figure*}[tbh]
      \begin{center}
      \includegraphics[width=0.8\textwidth,trim=0 0 0 0, clip]{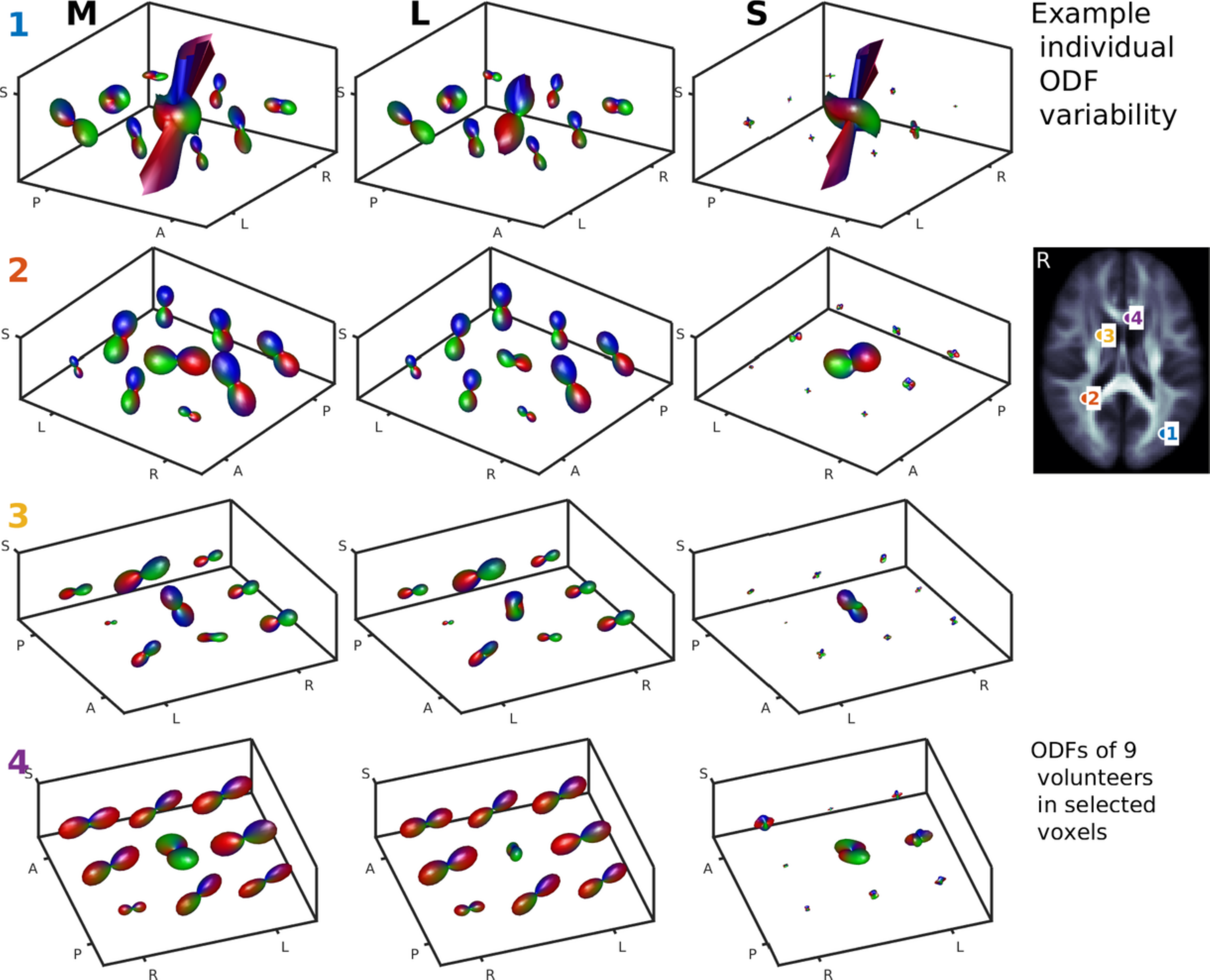}
          \caption{Examples of the separation of $M$ in ODF-features ($L$) and individual variability ($S$) using the ODF $L+S$ approach in a cohort of healthy HCP volunteers. Each row shows an ODF in the indicated voxel with a large variability (central ODF) relative to the ODFs of the preceding and succeeding volunteers in that voxel (8 surrounding ODFs). ODFs from $M$, $L$ and $S$ are displayed in the left, middle and right column respectively.\label{FigOutlierODF}}
      \end{center}
  \end{figure*}


\end{document}